\documentclass[prb,twocolumn,a4paper,superscriptaddress,showpacs]{revtex4}

\usepackage{graphicx}
\usepackage{psfrag}
\usepackage{color}
\usepackage[below]{placeins}
\usepackage{amsmath}
\graphicspath{{Bilder/}}

\def\be{\begin{equation}}             \def\ee{\end{equation}}
\def\bea{\jot0ex\begin{eqnarray}}     \def\eea{\end{eqnarray}}
\def\re#1{\ref{#1}}
\def\ie{{\it i.e.,~}}
\def\eg{{\it e.g.,~}}
\def\kago{kagom\'{e}}
\def\SPN{{\rm Sp(\CN)}}
\def\SP#1{{\rm Sp(#1)}}
\def\SUZ{{\rm SU(2)}}

\def\CN{\mathcal{N}}
\def\ND{N_{\bigtriangledown}}

\def\CH{\mathcal{H}}
\def\CHt{\mathcal{H}_{\bigtriangleup}}
\def\EMF{E_{\rm MF}}
\def\HMF{\mathcal{H}_{\rm MF}}
\def\HAKAF{\mathcal{H}_{\rm AKAF}}
\def\la{\lambda_a} \def\lc{\lambda_c}
\def\del{\mbox{\boldmath{$\delta$}}}
\def\fdel{\mbox{\boldmath{\scriptsize$\delta$}}}
\def\feps{\mbox{\boldmath{$\varepsilon$}}}
\def\k{{\mathbf{k}}}
\def\q{{\mathbf{q}}}
\def\kmin{{\mathbf{k}_{min}}}
\def\qord{{\mathbf{q}_{ord}}}

\def\S{{\mathbf{S}}}
\def\xm{\xi_{min}(\alpha,\beta)}
\def\wf{w_F(\k)}
\def\wdc{w_{DC}(\k)}
\newcommand{\bra}[1]{{\left< #1 \right|}}
\newcommand{\ket}[1]{{\left| #1 \right>}}
\def\NL#1{\nonumber \\[#1]}
\def\Co{(Color online) }

\begin{document}

\title{Heisenberg antiferromagnet with anisotropic exchange 
on the kagom\'{e} lattice:
Description of the magnetic properties of volborthite} 

\author{T. Yavors'kii}
\affiliation{Department of Physics and Astronomy, University of Waterloo, 
200 University Avenue W, Waterloo, N2L 3G1, Canada.} 
\author{W. Apel}
\affiliation{Physikalisch-Technische Bundesanstalt,
Bundesallee 100, D-38116 Braunschweig, Germany.}
\author{H.-U. Everts }
\affiliation{Institut f\"{u}r Theoretische Physik, Leibniz Universit\"{a}t 
Hannover, Appelstra{\ss}e 2, D-30167 Hannover, Germany.} 

\date{\today}

\begin{abstract}
We study the properties of the Heisenberg antiferromagnet with spatially 
anisotropic nearest-neighbour exchange couplings on the {\kago} net, 
\ie with coupling $J$ in one lattice direction and couplings  $J'$ along 
the other two directions.
For $J/J'\gtrsim 1$, this model is believed to describe the magnetic properties of 
the mineral volborthite. 
In the classical limit, it exhibits two kinds of ground states: 
a ferrimagnetic state for $J/J' < 1/2$ and a large manifold of canted 
spin states for $J/J' > 1/2$.
To include quantum effects self-consistently, we investigate the $\SPN$ 
symmetric generalisation of the original $\SUZ$ symmetric model in the 
large-$\CN$ limit. 
In addition to the dependence on the anisotropy, the $\SPN$ symmetric model 
depends on a parameter $\kappa$ that measures the importance of quantum effects. 
Our numerical calculations reveal that in the $\kappa$-$J/J'$ plane, 
the system shows a rich phase diagram containing a ferrimagnetic phase, 
an incommensurate phase, and a decoupled chain phase, the latter two 
with short- and long-range order. 
We corroborate these results by showing that the boundaries between the 
various phases and several other features of the $\SPN$ phase diagram 
can be determined by analytical calculations. 
Finally, the application of  a block-spin perturbation expansion to the trimerised version 
of the original spin-$1/2$ model leads us to suggest that in the limit of strong 
anisotropy, $J/J' \gg 1$, the ground state of the original model is a collinearly 
ordered antiferromagnet, which is separated from the 
incommensurate state by a quantum phase transition.
\end{abstract}


\maketitle

\section{Introduction}
\begin{figure}
\begin{minipage}{9cm}
\includegraphics[height=3.4cm]{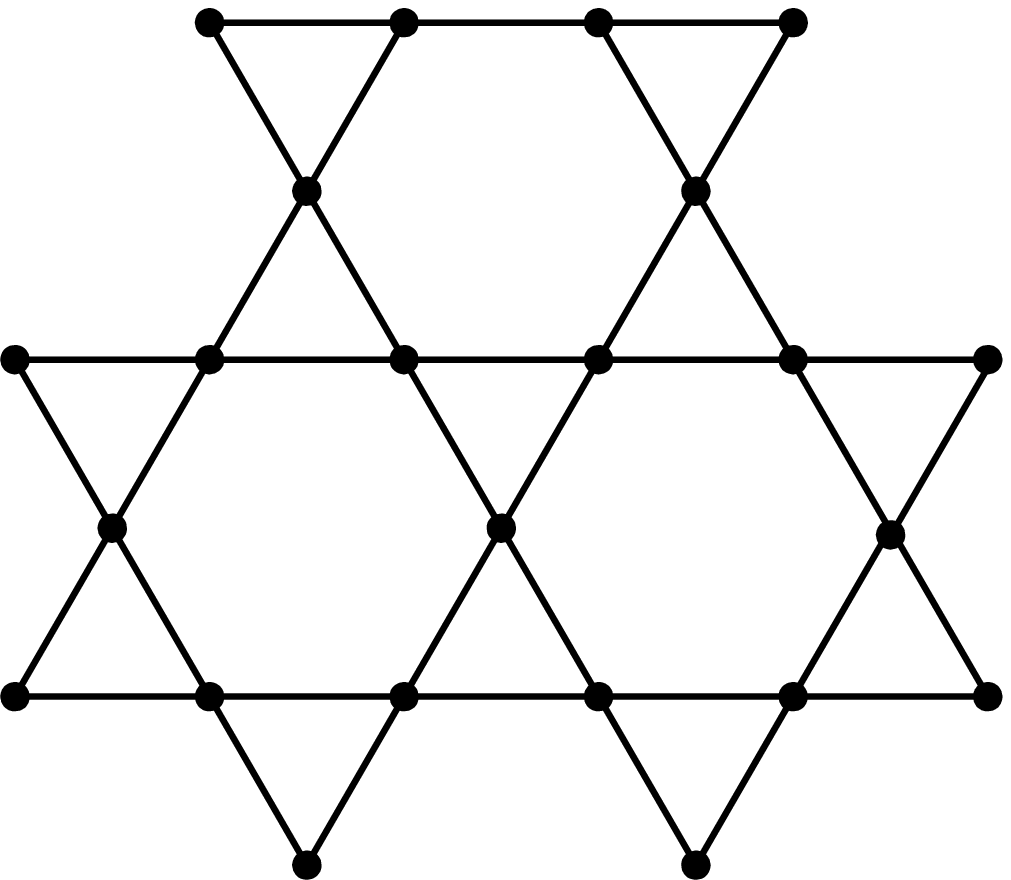}
\hspace{3mm}
\includegraphics[height=3.6cm]{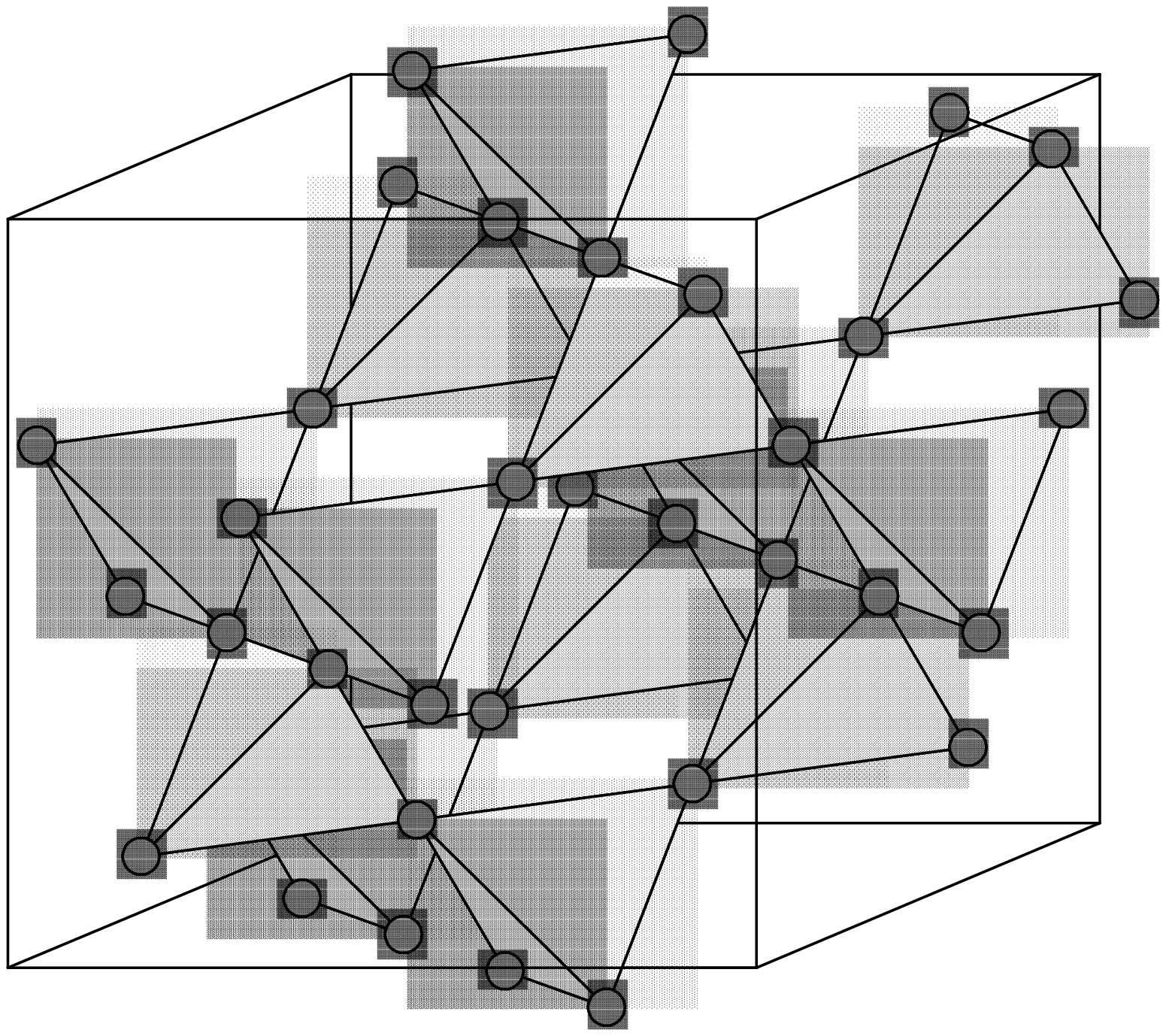}\\
\hspace{1.75cm}(a) \hspace{3.5cm} (b) \hspace{1.75cm}
\end{minipage}
\caption{{\kago} lattice (a), pyrochlore lattice (b)}
\label{iskago}
\end{figure}

In the ongoing search for novel states of condensed matter, frustrated 
antiferromagnets have played a key role (for a recent review, see 
Ref.~\onlinecite{ML04}). 
Among the many substances that have been investigated experimentally 
and the numerous spin models that have been studied theoretically, 
those in which the magnetic ions occupy the vertices of corner-sharing 
frustrating entities have attracted particular attention in this context. 
The best known examples are the {\kago} antiferromagnet (KAF), consisting 
of corner sharing triangles, and the pyrochlore antiferromagnet, 
consisting of corner sharing tetrahedra (see Fig.~\ref{iskago}). 

\begin{figure}
\psfrag{a}{$a$} \psfrag{b}{$b$} \psfrag{c}{$c$}
\psfrag{b1}[bc][Bl]{$[i,j]$}
\psfrag{b2}[cc][Bl]{{\rotatebox{-60}
  {\rule{0pt}{4mm}$\langle i, k \rangle$}}}
\psfrag{b3}[cc][Bc]{{\rotatebox{60}{$\langle k, j \rangle$}}}
\psfrag{J}{\color{red}{$J$}} \psfrag{Js}{$J'$}
\psfrag{d1}{$\del_1$} \psfrag{d2}{$\del_2$} \psfrag{d3}{$\del_3$}
\includegraphics[width=8cm]{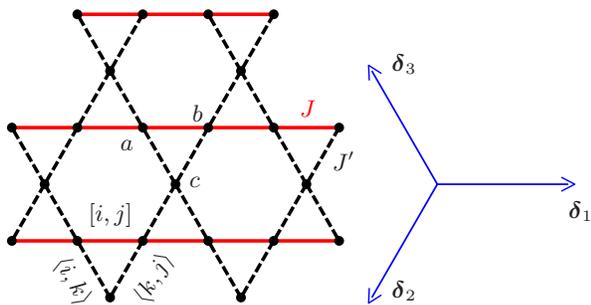}
\caption{\Co Anisotropic {\kago} model. The coupling $J'$ and the nearest 
neighbour distance will be set equal to unity in the calculations.
$\del_1 (\parallel \hat{e}_x)$, $\del_2$, and $\del_3$ are the three 
primitive lattice vectors of the {\kago} net.}
\label{aniskago}
\end{figure}

The main distinction between the KAF, the pyrochlore antiferromagnet and 
other frustrated and unfrustrated magnets is the large ground-state 
degeneracy of the former:
classical Heisenberg antiferromagnets with nearest-neighbour interactions 
on corner-sharing lattices have a large ground-state degeneracy,
which in the above two examples even leads to a finite ground-state 
entropy (see, \eg Ref.~\onlinecite{M01} and references therein). 
Quantum effects may lift this degeneracy, and, indeed, in numerical 
studies of small cells of the spin $\frac{1}{2}$ KAF, an exponentially 
large number of very low-lying quantum states has been observed 
\cite{LBLPS97, WEBLSLP98}. 
It has been suggested that this abundance of low-lying states can be 
understood in a description of the low-energy physics of the quantum KAF 
as spin liquid consisting of nearest-neighbour spin singlets 
\cite{M98,MM00}. 
However, a complete picture of the ground state and of the excitations of 
the KAF is still missing.
Further theoretical, but also experimental studies with emphasis on the 
quantum properties of the KAF are therefore highly desirable. 
In this last respect, the mineral volborthite is a very promising 
candidate. 
It has been the subject of several recent experimental investigations 
\cite{HHKNTKT01,FFGILSUKLRKKHH03,BBMTMABH04,BBMLDTM05}. 
The magnetic lattice of this natural antiferromagnet consists of the 
$S=1/2$ spins of $Cu^{2+}$ ions that are located on the vertices of well 
separated planar {\kago}-like nets. 
A monoclinic distortion of the lattice leads to a slight difference 
between the exchange couplings along one lattice direction ($J$) and the 
two other directions ($J'$)(see Fig.~\ref{aniskago}).   
Since neither signs of long-range order nor signs of a spin-gapped 
singlet ground-state were found in experiments on volborthite, 
the substance seems to be a good candidate for the observation of the 
low-energy features that are thought to be typical for {\kago} type 
antiferromagnets \cite{ML04}.

Whether and to what extent the different exchange couplings along  
different lattice directions of the {\kago} net of volborthite 
influence the low-energy physics of the system is presently unknown. 
In the present paper, we study this question on the basis of the model 
Hamiltonian

\be
\HAKAF = J \sum_{[i,j]} \S_i \S_j + J' \sum_{\langle k,i \rangle} 
 \S_k \S_i\;.
\label{akago}
\ee

The symbols $[i,j]$ and $\langle k,i \rangle$ denote, respectively, bonds 
between nearest-neighbour sites on the horizontal chains ($a$, $b$) 
and  bonds between the middle sites ($c$) and the  sites $a$, $b$, 
see Fig.~\re{aniskago}. 
Since the physics of this model depends only on the ratio $J/J'$ of the 
exchange constants, we set $J' = 1$ in the sequel.
We will consider the spatially anisotropic {\kago} antiferromagnet (AKAF), 
Eq.~(\re{akago}), in the full range of $J$, $0 < J < \infty$ since 
this is of theoretical interest: 
one expects to see quantum phase transitions as $J$ is increased.
It is of particular interest to find out whether there is a transition 
from two-dimensional magnetic states to a set of decoupled chains 
with free spins on the axes between the chains for large values of $J$.

The paper is organised as follows. 
In Sec.~\ref{classical}, we consider the model (\ref{akago}) in the 
classical limit. 
At this level, we find no sign of a transition from the two-dimensional 
magnet to a set of decoupled chains as $J$ increases to infinity. 
Nonetheless, the ground-state degeneracy, as well as the spin wave 
spectrum are found to change qualitatively as the anisotropy of the 
model varies.
In Sec.~\ref{SPN}, we consider a generalisation of the $\SUZ$ symmetric  
model (\re{akago}) to the $\SPN$ symmetric version \cite{SR91,S92}
and describe its properties in the large-$\CN$ limit, where a mean-field 
treatment of the model is adequate.
We obtain a detailed description of how possible ground states  
of the model depend on the coupling $J$ and on the spin length $S$. 
A fairly rich phase diagram with a ferrimagnetic phase for small $J$, 
long-range ordered and short-ranged incommensurate phases for intermediate 
values of $J$, and a decoupled-chain phase for large $J$ emerges. 
Parts of these results have been published previously, see 
Ref.~\onlinecite{AYE07}.
In Sec.~\ref{quantum}, we devise trial quantum ground states of the 
original $S=1/2$ model.
We chose the states such that they are exact eigenstates of 
$\HAKAF$, if the couplings on the upward pointing triangles of 
Fig.~\ref{aniskago} are switched off, and we then treat these 
couplings perturbatively. 
In the limit $J\longrightarrow \infty$ 
this yields an effective Hamiltonian for the spins on the $c$ sites 
which represents an anisotropic triangular antiferromagnet. 
The conclusions of Starykh and Balents \cite{SB07} about the ground state 
of this effective model lead us to conjecture the existence of
a quantum phase transition in the AKAF for large $J$.  
In Sec.~\ref{summary}, we summarise and discuss our results. 
In two Appendices, we present technical details of the counting procedure 
for the classical ground-states, and of the Ginzburg-Landau type procedure 
that allows us to determine the boundaries in the phase diagram 
analytically.

\section{Classical and semiclassical aspects}
\label{classical}
Similar to other isotropic spin models on lattices with triangular 
elementary cells, the {\em classical} ground states of $\HAKAF$, 
Eq.~(\re{akago}), are spin configurations, which satisfy the condition 
that for each elementary triangular plaquette of the lattice, 
Fig.~\ref{aniskago}, the energy is minimal. 

\begin{figure}[b]
\includegraphics[width=4cm]{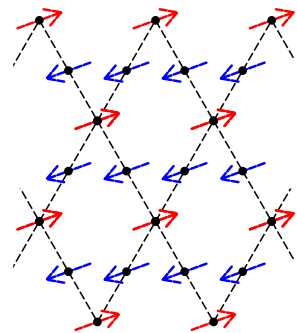} 
\caption{\Co Ferrimagnetic state for $J = 0$, \ie when there is no coupling 
between chain spins, cf.~Fig.~\ref{aniskago}.}
\label{FerriKago}
\end{figure}

\begin{figure}[t]
\psfrag{qx}[r][r]{\Huge $q_x$}
\psfrag{qy}[l][l]{\Huge $q_y$}
\psfrag{om}[r][r]{\Huge $\omega_{\alpha}(\q)$}
\includegraphics[angle=-90,width=7cm]{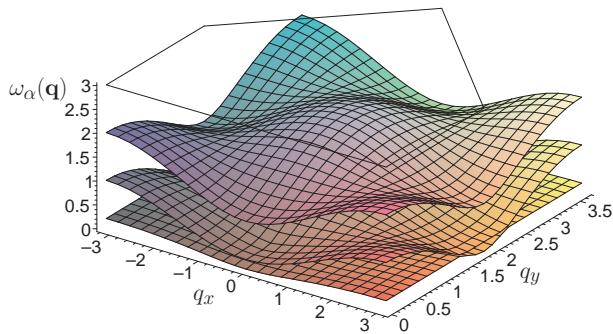}
\caption{\Co Spin-wave frequencies $\omega_{\alpha}(\q)$, $\alpha=1,2,3$
for $J=0.4$; the contour at the top of the plot marks half the Brillouin zone. }
\label{fsw04}
\end{figure}

For $J = 0$, this yields a ferrimagnetic state with the chain spins 
aligned in one direction and the middle spins pointing in the opposite 
direction, so that the total magnetisation is $M = \ND S$
($\ND $: number of downward pointing triangles, $\ND = N_s/3 $ where $N_s$ 
is the number of sites of the system). 
We illustrate this situation in Fig.~\ref{FerriKago}. 
According to the Lieb-Mattis theorem, the exact {\em quantum} ground state 
(GS) of the model $\HAKAF$ also has total spin $S^{tot} = \ND\,S$ for 
$J = 0$, see Ref.~\onlinecite{LM62}, \ie for $J = 0$, the quantum GS 
is ferrimagnetic too. 
By continuity, one expects the quantum GS to remain ferrimagnetic for 
sufficiently small finite $J$. 
This will be confirmed by our considerations of the large-$\CN$ limit 
of the $\SPN$ version of our model (see the analytical and numerical 
work in Sects.~\ref{SPN}, \ref{results} and Appendix \ref{dc}) 
and by the block spin perturbation approach (Sec.~\ref{quantum}). 
{\em Classically}, the ferrimagnetic state remains stable up to $J = 1/2$. 
The excitation spectrum of the ferrimagnetic state obtained in linear 
spin-wave (LSW) approximation is shown in Fig.~\ref{fsw04}. 

The analytic expressions for these three frequency surfaces 
$\omega_{\alpha}(\q)$, $\alpha=1,2,3$, are obtained as solutions of a 
third-order secular equation and are too lengthy to be presented  here. 
However, one can easily assure oneself that the dispersion of the 
gapless mode is quadratic at the origin. 
Thus, one has the typical mode structure of a ferrimagnet here with one 
ferromagnetic mode and two optical modes, see, \eg Ref.~\onlinecite{BMY97}.
As $J$ increases towards $1/2$, the ferromagnetic frequency surface 
looses its dispersion and turns into a plane of zero modes, one zero mode 
for each wave vector in the magnetic Brillouin zone (BZ), at $J = 1/2$. 
The gap of the lower optical mode closes at this value of $J$ in the 
centre of the BZ and the dispersion of this mode becomes linear 
for small wave vectors as for an antiferromagnetic spin-wave mode.  
         
\begin{figure}[b]
\includegraphics[width=2.5cm]{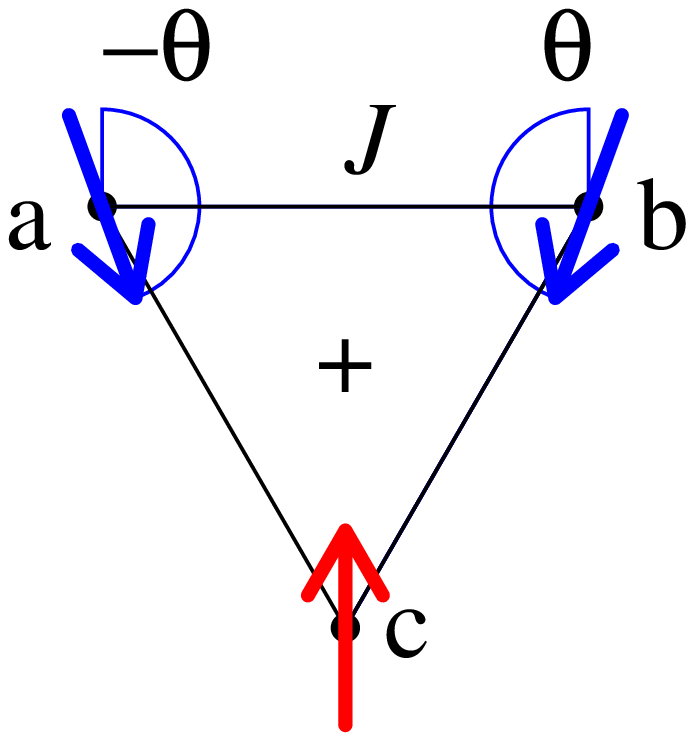} \hspace{1.5cm}
\includegraphics[width=2.5cm]{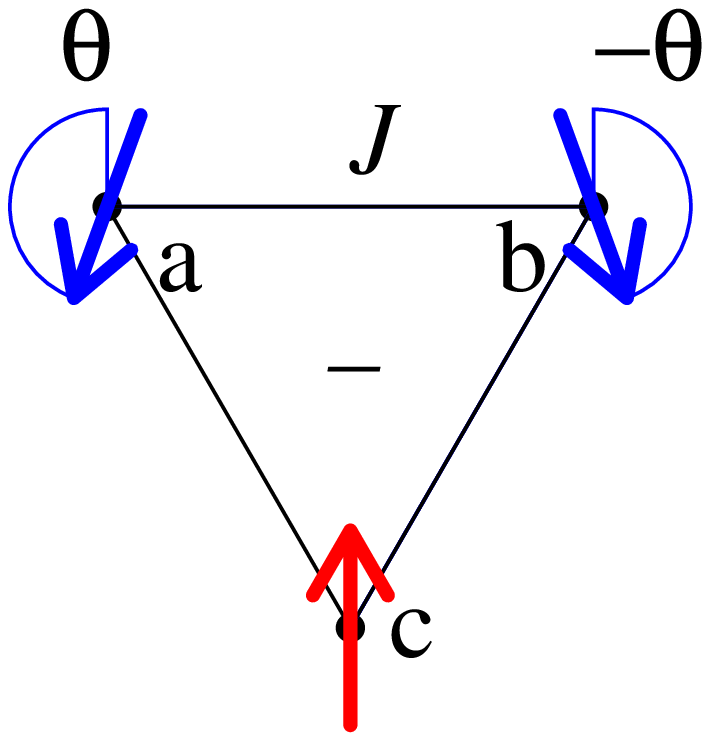}
\caption{\Co Canted spins of the AKAF at $J>1/2$.}
\label{cansp}
\end{figure}
\begin{figure}[t]
\psfrag{qx}[r][r]{\Huge $q_x$}
\psfrag{qy}[l][l]{\Huge $q_y$}
\psfrag{om}[r][r]{\Huge $\omega_{\alpha}(\q)$}
\includegraphics[angle=-90,width=7cm]{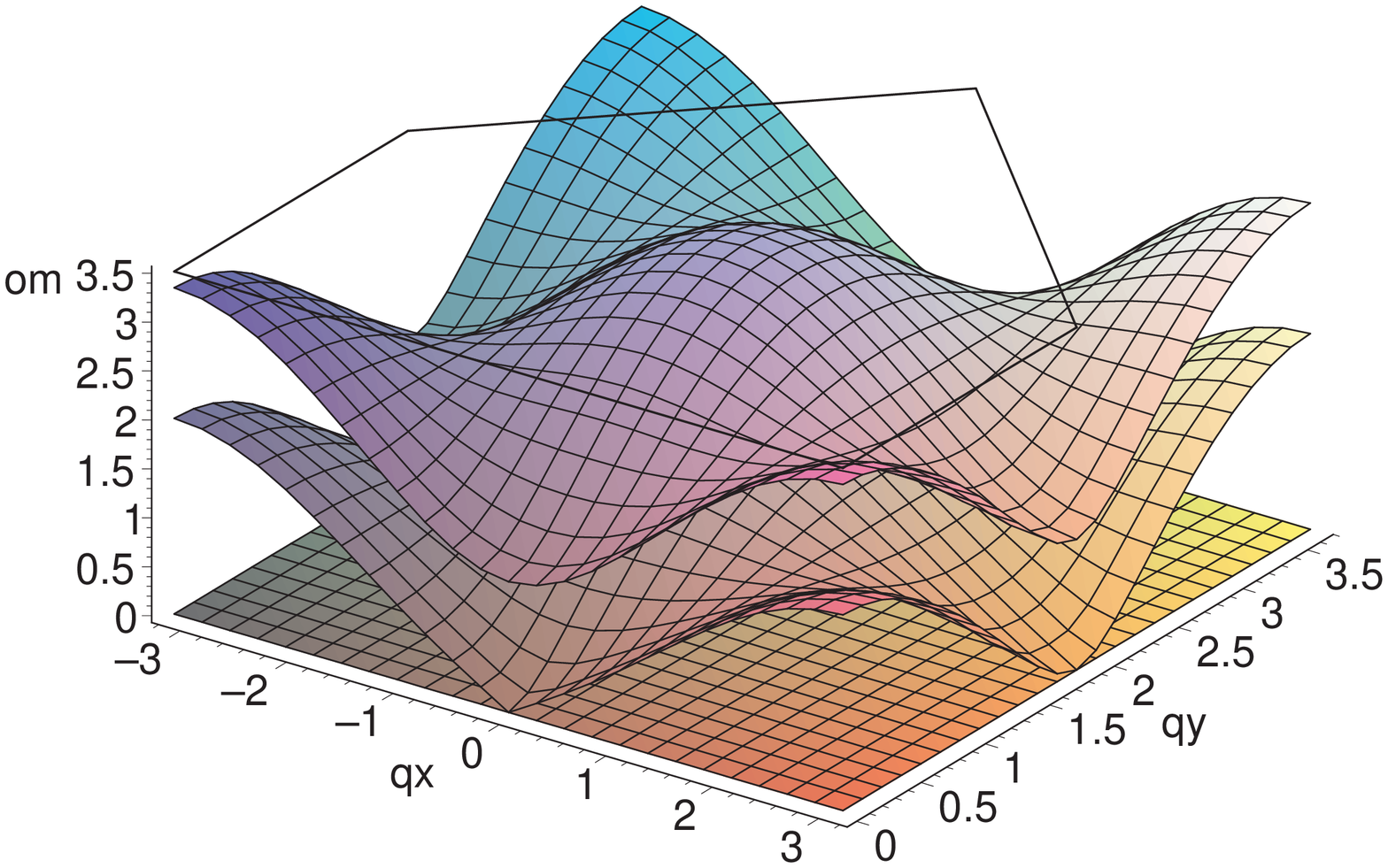}
\caption{\Co Same as Fig.\re{fsw04} for $J=0.6$.}
\label{sw06}
\end{figure}
\begin{figure}[b]
\psfrag{qx}[r][r]{\Huge $q_x$}
\psfrag{qy}[l][l]{\Huge $q_y$}
\psfrag{om}[r][r]{\Huge $\omega_{\alpha}(\q)$}
\includegraphics[angle=-90,width=7cm]{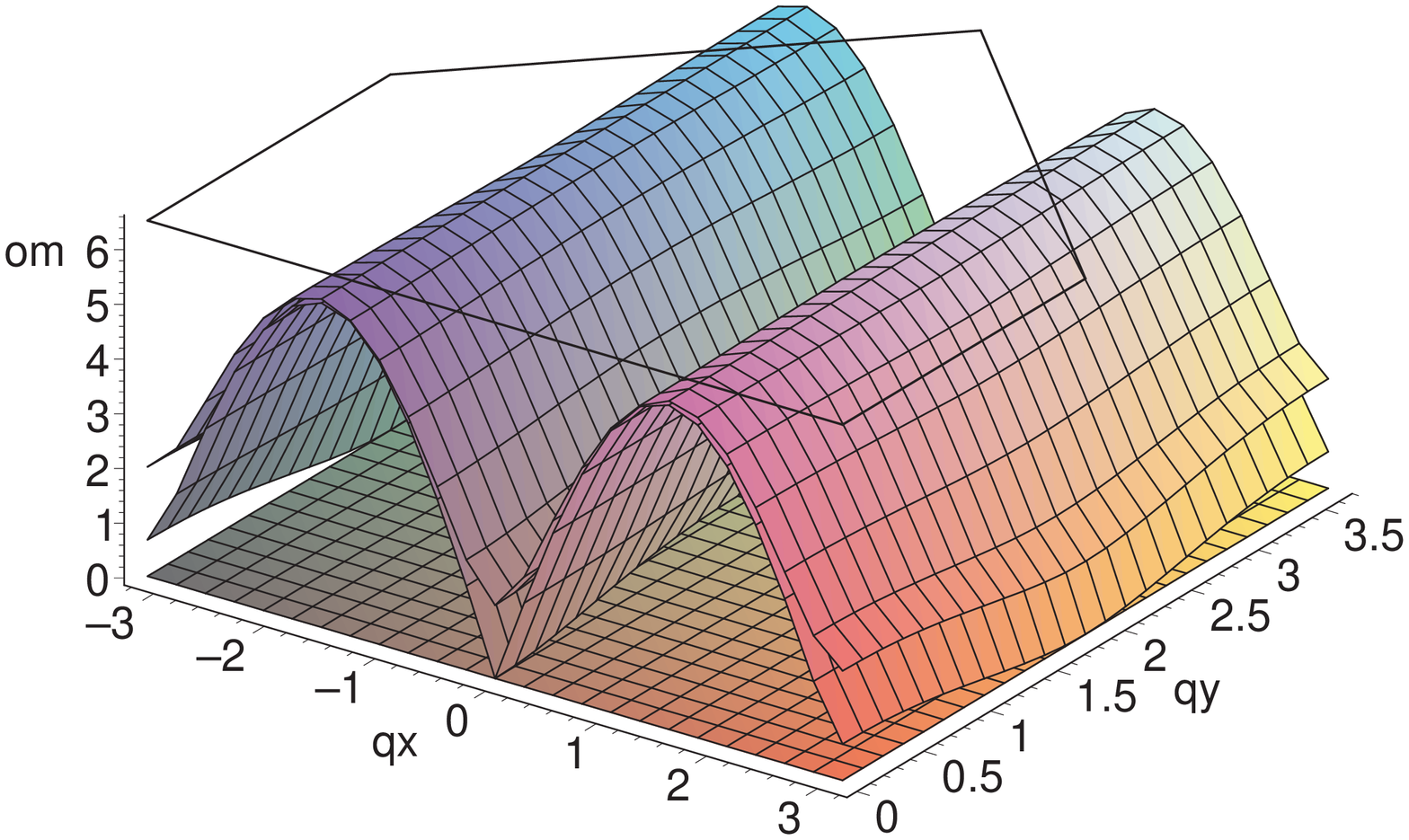}
\caption{\Co Same as Fig.\re{fsw04} for $J=3$.}
\label{sw3}
\end{figure}

At $J = 1/2$, the classical GS configuration changes from 
the unique ferrimagnetic state to an ensemble of degenerate canted 
{\em coplanar} states. 
These states are characterised by two variables: the angle $\theta$,  
which the middle spin of a given triangular plaquette forms with the two 
chain spins of the same plaquette (see Fig.~\ref{cansp}), and the two 
valued chirality $\chi = \pm 1$, which denotes the direction in which 
the spins turn as one moves around the plaquette in the mathematically 
positive sense.

For $J \geq 1/2 $, the requirement that the energy of any of the 
elementary triangular plaquettes of the lattice Fig.~\ref{aniskago} 
be minimal is $\theta= \arccos(-1/(2J))$, ($\theta > 0$).  
The different degenerate canted states arise from different possibilities 
to assign positive or negative chiralities to the pla\-quettes of the 
lattice. 
We show in the Appendix \re{gsd} that for the general case of 
$\theta \neq 2\pi/3$ ($J \neq 1$), the number of spin configurations, 
$N^{aniso}_{GS}$ does not grow exponentially with the number of sites. 
Rather, $N^{aniso}_{GS} < 2^{\alpha \sqrt{\ND}}$, where $\alpha < 3$. 
This implies that the ground-state entropy per spin of the classical
AKAF vanishes in the thermodynamic limit. 
In this respect, the anisotropic model differs qualitatively from the 
isotropic KAF in the classical limit, which has an extensive entropy 
per spin. 
In the limit $J \rightarrow 1$, the anisotropic model approaches the 
isotropic KAF. 
Hence, one expects that for the anisotropic model there is an extensive 
number of low-lying excited states that become degenerate with the 
GS in the isotropic limit.

As in the case of the isotropic KAF, the spin-wave Hamiltonian 
is in linear order independent of the particular classical GS 
that has been chosen as the starting point of the expansion, 
Ref.~\onlinecite{HKB92}. 
This implies that lowest-order quantum fluctuation do not select one or a 
group of classical GSs as true GSs, \ie the possible 
ordering effects of quantum fluctuations are not captured by the linear 
spin-wave (LSW) approximation. 
Figs.~\ref{sw06}, \ref{sw3} show the spin-wave frequency surfaces for 
$J = 0.6$ and for $J = 3$. 
It is easy to show analytically that, as is illustrated in these figures, 
the plane of zero frequency modes persists for all values of $J$ greater 
than $1/2$.  
The surfaces for $J < 1/2$ and for $J > 1/2$ join smoothly at $J = 1/2$. 
Thus, in the LSW approximation, the transition from the 
ferrimagnetically ordered state to the canted spin states appears to be of 
second order.
For $ J \gg 1$, the nonzero frequencies gradually loose their dispersion 
perpendicular to the strong-$J$ direction and take the shape of the 
spin-wave spectrum of antiferromagnetic chains parallel to this direction. 
However, no sign of a further transition from the canted spin states 
to a set of decoupled spin chains is found in this semiclassical approach.
In the next section, we will consider the symplectic $\SPN$ generalisation
of the antiferromagnetic model $\HAKAF$ in the large-$\CN$ limit. 
This approach, which was first proposed by Read and Sachdev, 
Refs.~\onlinecite{SR91, S92}, as a method to study frustrated 
antiferromagnets, has the benefit of including the ordering effects 
of quantum fluctuations self-consistently. 
It is of particular interest for spin models with two or more competing 
exchange couplings in the different lattice directions or over different 
lattice distances such as the present model, the $J_1$-$J_2$-$J_3$ model 
\cite{SR91}, the Shastry-Sutherland antiferromagnet \cite{CMS01} and 
the anisotropic triangular antiferromagnet \cite{CMK01}. 
For these models, it has provided an unbiased selection of possible 
GSs that may or may not be ordered depending on the value of 
a parameter $\kappa$, which is connected with the spin length $S$ 
(see below).

\section{Mean field $\SPN$ approach}
\label{SPN}
\subsection{Brief review of the method}

For a general antiferromagnetic Heisenberg model with a positive 
interaction matrix $J_{ij}$,
\be
  \mathcal{H} = \sum_{i>j} J_{ij} \; {\bf S}_i \cdot {\bf S}_j\,,
\ee

the $\SPN$ generalisation reads 

\be
\mathcal{H}_{\SPN} = - \sum_{i>j} \frac {J_{ij}}{2\CN} 
(\mathcal{J}^{\alpha \beta} b^{\dagger}_{i\alpha}  b^{\dagger}_{j\beta})
(\mathcal{J}_{\gamma \delta} b_i^{\gamma}  b_j^{\delta})\,. 
\label{HN}
\ee
Here, 
\be
\mathcal{J} = \left(\begin{array}{lll}
   \feps & & \\ & \feps & \\ & & \ddots 
\end{array} \right)
\ee
is the $2\CN \times 2\CN$ generalisation of the $2 \times 2$ antisymmetric 
tensor 
\be
\feps = \left(\begin{array}{c@{\hspace{1mm}}c} 0 & +1\\ -1 & 0 
        \end{array} \right)\,,
\ee
and $b_i^{\alpha}$ with $\alpha = 1, \dots, 2\CN$ are the $\SPN$ boson 
annihilation operators. 
(Here and in the sequel, we closely follow the notation of 
Ref.~\onlinecite{S92}; in particular, summation over repeated upper and 
lower indices is implied.)  
Thus, 
$\mathcal{J}^{\alpha \beta} b^{\dagger}_{i\alpha}  b^{\dagger}_{j \beta}$ 
is the generalisation of the creation operator
 $\varepsilon^{\alpha \beta} b^{\dagger}_{i\alpha} b^{\dagger}_{j\beta}$ 
 for a singlet on the bond $(i,j)$.
For the special case $\CN = 1$, one finds 
\be
(\mathcal{J}^{\alpha \beta} b^{\dagger}_{i\alpha}
  b^{\dagger}_{j\beta})
(\mathcal{J}_{\gamma \delta} b_i^{\gamma}  b_j^{\delta}) = 
-2{\mathbf S_i} \cdot {\mathbf S_j} + n_{bi}n_{bj}/2 + \delta_{ij}  n_{bi}
\,,
\ee
where 
\be
n_{bi} = b^{\dagger}_{i\alpha} b_i^{\alpha}
\label{NBOS}
\ee
is the boson number operator at site $i$ and where
\be
{\mathbf S_i} = b^{\dagger}_{i\alpha}  
  {\mbox{\boldmath{$\tau$}}}^{\alpha}_{\beta} b_{i}^{\beta}/2
\label{S}
\ee 
is the usual $\SUZ$ spin operator at site $i$. 
(${\mbox{\boldmath{$\tau$}}}$ are the Pauli matrices).  
Then, if one imposes the constraint that the number of bosons is the same 
for all lattice sites, $n_{bi} \equiv n_{b}$, the Hamiltonian 
$\mathcal{H}_{\SP{1}}$ is the familiar $\SUZ$ invariant antiferromagnetic 
Heisenberg Hamiltonian (plus some constants) with $n_b = 2 S$. 

In the subsequent exposition, we shall consider a Hamiltonian of the form 
(\re{HN}) in the large-$\CN$ limit following the strategy of 
Refs.~\onlinecite{SR91, S92}. 
Depending on the values of the couplings $J_{ij}$ and of $\kappa$, the GS 
of $\mathcal{H}_{\SPN}$ may either break the global $\SPN$ symmetry and 
exhibit LRO or it may be  $\SPN$ symmetric with only SRO.
Breaking of the $\SPN$ symmetry will happen through condensation, 
\ie by macroscopic occupation of one of the Bose fields $b_{\alpha}$. 
To allow for this, we introduce the parametrisation

\be
b_{i}^{m \sigma} = \left(
  \begin{array}{l} \sqrt{\CN} x_i^{\sigma}\\ \tilde{b}_i^{\tilde{m}\sigma}
  \end{array} \right)
\label{XB}
\ee

with $\alpha =(m\sigma)$, $m = 1, \cdots, \CN$, 
$\tilde{m} = 2, \cdots, \CN$ and $\sigma = \uparrow, \downarrow$.  
The field $x_{i}^{\sigma}$ is proportional to the condensate amplitude, 
$\langle b_i^{m\sigma} \rangle = \sqrt{\CN}\;\delta_1^{m}\;x_i^{\sigma}$. 
Aiming at a mean field treatment of the Hamiltonian $\mathcal{H}_{\SPN}$, 
which becomes exact in the large  $\CN$ limit, we decouple the quartic part 
by the Hubbard-Stratonovich technique with complex fields $Q_{ij} = - Q_{ji}$
and with Lagrange multipliers $\lambda_i$ that enforce the local constraints (\re{NBOS}).
The variables $Q_{ij}$ which are defined on nearest neighbour bonds of the lattice
are expectation values of the bond singlet creation 
operators in the GS, 
$Q_{ij} = \langle \sum_{\sigma \sigma'} \varepsilon^{\sigma \sigma'}
b_{i\,m\sigma}^{\dagger} b_{j\,m\sigma'}^{\dagger} \rangle$
and are to be determined self-consistently from the mean field type Hamiltonian

\begin{widetext}
\be
\HMF =  \sum_{i>j} \left\{ \frac{\CN}{2} J_{ij}|Q_{ij}|^2 
- \frac{1}{2} J_{ij} \left[ Q_{ij} \varepsilon_{\sigma \sigma'} 
\left(\CN x_i^{\sigma} x_j^{\sigma'}  + 
\sum_{\tilde{m}} \tilde{b}_i^{\tilde{m}\sigma}  
                 \tilde{b}_j^{\tilde{m}\sigma'} 
\right) +  h.c. \right] \right\} 
+ {} \sum_i \lambda_i \left(\CN |x_i^{\sigma}|^2 + 
\sum_{\tilde{m}} \tilde{b}^{\dagger}_{i\tilde{m}\sigma} 
      \tilde{b}_i^{\tilde{m} \sigma} - n_b \right) \;.
\label{HMF}
\ee  
\end{widetext}

The variational ground state energy, $\EMF$, of $\HMF$ in the large-$\CN$ 
limit is obtained by diagonalising the bosonic part of $\HMF$, 
integrating over the $2(\CN - 1)N_s$ bosonic fields 
$\tilde{b}_i^{\tilde{m}\sigma}$ in the action associated with $\HMF$.
One obtains:
\bea
\frac{\EMF}{\CN} &=& \sum_{i>j} \left[\frac{1}{2} J_{ij} |Q_{ij}|^2 
- \frac{1}{2} J_{ij} \left(Q_{ij} \varepsilon_{\sigma \sigma'} 
  x_i^{\sigma}  x_j^{\sigma'} + h.c. \right) \right] \nonumber \\[2mm]
&&+ \sum_{\k,\mu}\omega_{\mu}(\k;Q, \lambda)
  + \sum_i \lambda_i \left(|x_i^{\sigma}|^2 - 1 - \kappa\right) \,.  
\label{EMF}
\eea
Here $\omega_{\mu}(\k;Q, \lambda)$ are the positive eigenvalues 
of the bosonic part of $\HMF$, and $\kappa=n_b/\CN$ is kept fixed 
in the limiting procedure \cite{SR91,S92}. 
The parameter $\kappa$ is a measure for the importance of quantum 
fluctuations:
by varying $\kappa$ from small to large values, one drives the system from 
the regime dominated by quantum fluctuations to the classical regime, 
\ie from the disordered into the ordered region.
Finally, the GS is obtained by finding the saddlepoint of $\EMF$
in the space of the variables $Q_{ij}$ and $x^{\sigma}_i$ subject to 
the constraints

\be
\partial \EMF(Q, \lambda)/\partial \lambda_i = 0\,.  \label{DIFFCON}
\ee

In addition to the GS itself, the spin-spin correlation function 
$G_{ij} = \langle {\mathbf S_i} \cdot {\mathbf S_j} \rangle$ in the GS 
is an important piece of information. 
In particular, by considering its behaviour in the limit 
$|i - j| \longrightarrow \infty$, one can distinguish between LRO 
and SRO. 
According to Sachdev \cite{S92}, to obtain $G_{ij}$ 
in the $\SPN$ symmetric approach, the $\SUZ$ invariant expression 
${\mathbf S_i} \cdot {\mathbf S_j}$ must be replaced by the $\SPN$ 
invariant expression
\be 
\frac{1}{4\CN^2}(b_{i \alpha}^{\dagger} b_i^{\beta} b_{j \beta}^{\dagger}  b_j^{\alpha} 
- \mathcal{J}^{\alpha \gamma} \mathcal{J}_{\beta \delta} 
b_{i \alpha}^{\dagger}  b_i^{\beta} b_{j \gamma}^{\dagger}b_j^{\delta}).
\label{SPNCORR}
\ee

Within the mean field approach, $G_{ij}$ can then be calculated
straightforwardly.

\subsection{The anisotropic {\kago} antiferromagnet}
\subsubsection{Choice of mean-field variables}
We wish to apply the procedure described above to the AKAF
represented by the Hamiltonian (\re{akago}). 
To render the problem of finding the eigenvalues $\omega_{\mu}$ in 
Eq.~(\re{HMF}) and of optimising $\EMF$ tractable, we have to restrict 
the number of variables $Q_{ij}$ and $\lambda_i$. 
We do so by demanding that the mean field Hamiltonian $\HMF$ for the 
spinon operators $b^{(\dagger)}$ is symmetric under transformations of 
the projective symmetry group (PSG) that is related to the symmetry group 
of the spin Hamiltonian $\HAKAF$ (Eq.~\re{akago}) (see Ref.~\onlinecite{WV06}).
We include two translations, a rotation by $\pi$ and a mirror axis orthogonal
to the preferred direction of the exchange constants ($J$).
Thus generalising the treatment of Wang and Vishwanath to our model, we find 
eight mean-field states with different symmetries.
Seven of them have flux in the sense of Ref.~\onlinecite{TMS06} in 
various cells of the lattice.
Following the arguments in Ref.~\onlinecite{TMS06}, we exclude all 
flux-carrying states and end up with the solution (cf.~Fig.~\re{QP})
$P_{1,2,3} = Q_{1,2,3}$, $Q_3 = Q_2$, and $\lambda_b = \lambda_a$.

\begin{figure}[h]
\psfrag{Q1}[r][r]{\small $Q_1$}
\psfrag{Q2}[r][r]{\small $Q_2$}
\psfrag{Q3}[r][r]{\small $Q_3$}
\psfrag{P1}[r][r]{\small $P_1$}
\psfrag{P2}[r][r]{\small $P_2$}
\psfrag{P3}[r][r]{\small $P_3$}
\psfrag{a}{$a$} \psfrag{as}{$a'$}
\psfrag{b}{$b$} \psfrag{bs}{$b'$}
\psfrag{c}{$c$} \psfrag{cs}{$c'$}
\includegraphics[width=8cm]{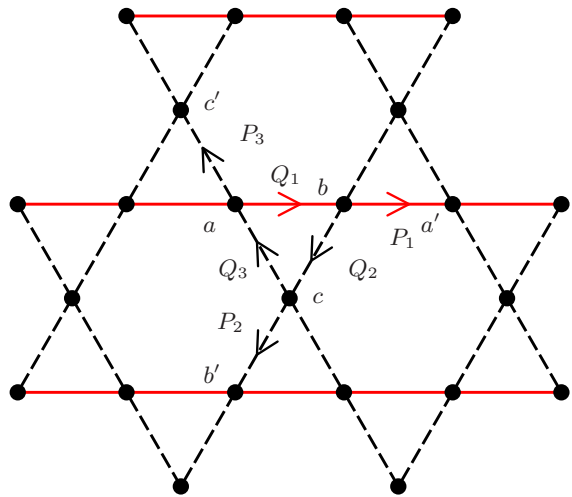}
\caption{\Co Arrangement of mean field parameters: $Q_1 \equiv Q_{ab}$,
$Q_2 \equiv Q_{bc}$, and $Q_3 \equiv Q_{ca}$ denote the intra triangle bonds, 
$P_1 \equiv Q_{ba'}$, $P_2 \equiv Q_{cb'}$ and $P_3 \equiv Q_{ac'}$ denote 
the inter triangle bonds.  
$\lambda_a$, $\lambda_b$, and $\lambda_c$ are the Lagrange multipliers 
needed to implement the constraints on the sites $a$, $b$, and $c$. }
\label{QP}
\end{figure}

In order to check the flux-argument in Ref.~\onlinecite{TMS06}, we have 
explicitely studied the solution $P_{1,2,3} = -Q_{1,2,3}$ and found that 
it is always of higher energy (For $J=1$, this agrees with the result 
of Ref.~\onlinecite{S92}).

Thus, the expression Eq.~(\re{EMF}) can now be cast into the form
\setlength{\arraycolsep}{0mm}
\bea
\frac{\EMF}{\CN \ND} \;=\;
&& J\, |Q_1|^2 + 2\, |Q_2|^2 - (2 \la + \lc)(\kappa +1) \nonumber \\
&& + \frac{1}{\ND} \sum_{\k,\mu}\omega_{\mu}(\mathbf{k}) 
  \left(1 + |{x}_{\mu}(\mathbf{k})|^2 \right)\,, \label{EMFK}
\eea
where the condensate is written in diagonalized form and 
$\omega_{\mu}(\mathbf{k})$ are the three positive solutions of
\be 
\det \mathbf{\hat{D}}(\omega) = 0\,. 
\label{DETD}
\ee
Here,
\be
\mathbf{\hat{D}}(\omega) = \left(\begin{array}{c@{\hspace*{3mm}}c}
\mathbf{\hat{\Lambda}}-\omega \mathbf{\hat{I}} & \mathbf{\hat{Q}} \\
\mathbf{\hat{Q}^{\dagger}} & \mathbf{\hat{\Lambda}}+\omega \mathbf{\hat{I}}
\end{array} \right)\,,
\label{DP}
\ee
with 
\bea
\mathbf{\hat{\Lambda}} \,&=&\, diag(\lambda_a,\,\lambda_c,\,\lambda_a)
 \, , \\[5mm]
\mathbf{\hat{Q}} \,&=&\, \left( \begin{array}{ccc}
0 & \tilde{Q}_2(\mathbf{k}) & - J \tilde{Q}_1(-\mathbf{k})\\
-\tilde{Q}_2(-\mathbf{k}) & 0 & \tilde{Q}_3(\mathbf{k})\\
J \tilde{Q}_1(\mathbf{k}) & -\tilde{Q}_3(-\mathbf{k}) & 0
\end{array}
\right)\, , \label{MP}
\eea
\be
\mbox{and} \quad \tilde{Q}_a(\mathbf{k})  =  \frac{1}{2} Q_a \left(
 e^{i {\fdel}_a \mathbf{k} /2} - e^{-i {\fdel}_a \mathbf{k} /2}
\right)\,,\quad a=1, 2, 3, \label{TILQ}
\ee
$\del_{1,2,3}$, see Fig.~\ref{aniskago}. 
 
\subsubsection{Technical details of the numerical extremalisation}
\label{minimax}

Determination of the ground state of the AKAF in the considered 
approximation has been reduced to minimization of the Eq.~(\ref{EMFK}) 
with respect to two variables $Q_1$ and $Q_2$, subject to the Lagrange 
constraints with respect to two parameters $\lambda_a$ and $\lambda_c$.
Being apparently trivial, the optimization procedure turns out to be 
quite involved technically. 

First, we find it crucial to consider at least two different 
chemical potentials.
Other than for the spatially isotropic KAF, $J=1$, we were not able 
to find a non-trivial solution if we used a single $\lambda$,  
$\lambda_a=\lambda_b=\lambda_c$.
If $\lambda_a$ and $\lambda_c$ are different, 
$[\mathbf{\hat{\Lambda}}, \mathbf{\hat{Q}}]\neq 0$, the Lagrange 
multipliers enter the expressions for the frequencies $\omega_{\mu}$ 
non-trivially, other than in the case of a global uniform chemical 
potential (cf. Ref.~\onlinecite{S92}).
In turn, the Lagrange constraints cannot be satisfied semi-analytically,
and require a numerical treatment.
Second, we choose to work directly in the {\em thermodynamic limit} 
$N_s\rightarrow\infty$ of the model (\ref{EMFK}) by performing 
a numerical self-adapting integration over the BZ. In this limit, the 
singularities can be integrated, and symmetry breaking is signalled by 
the appearance of a finite value of the condensate amplitude ${x_{\mu}} (\k)$ 
at a certain wavevector $\k = \q_{ord}$, which characterises the type of magnetic 
order. 
We mention here that the extremalisation of a mean-field energy of the type of  
Eq.~(\ref{EMFK}) can also be achieved by solving the pertinent stationarity 
conditions numerically for {\em finite} systems, {\ie} for {\em finite} $N_s$, see \eg
Ref.~\onlinecite{MBL98}. 
Then, the type of magnetic order has to be detected 
by calculating the structure factor. 
Third, we see that the Eq.~(\ref{EMFK}) has a minimum with respect to the 
physical bond parameters $Q_1$ and $Q_2$ only after the elimination of the 
chemical potentials. 
In the full $Q-\lambda$ space we face an extremalization problem. 

Technically, we find it convenient to use a polar coordinate 
parametrisation for the variables $Q_1,\, Q_2$ and 
$\lambda_a,\, \lambda_c$:
\be
Q_1 = Q \cos(\alpha), \qquad Q_2 = Q \sin(\alpha)  \label{Q}\,,
\ee
\be
\la = \Lambda \sin(\beta), \qquad \lc = \Lambda  \cos(\beta)\,.
\ee
We perform an optimization with respect to the variables 
$Q, \Lambda,\alpha,\beta$, as well as condensate densities 
${x}_{\mu}(\mathbf{k})$ in accord with the following algorithm 
($J$ and $\kappa$ are kept fixed).

\def\theparagraph{\roman{paragraph}}
\paragraph{}
We fix the angles $\alpha,\,\beta$ and the amplitude $Q$, and first 
exploit the stationarity condition for $\EMF$ with respect to 
$\Lambda$. 
It is convenient to write the corresponding equation in the following form:
\bea
&&\left[2\sin(\beta)+\cos(\beta)\right](\kappa+1) \nonumber \\[2mm]
&&\;\;- \frac{1}{\Omega} \int_{\rm B.Z.} \!\!d^2 k \; \sum_{\mu} 
        \left|{x}_{\mu}(\k)\right|^2 \;
        \partial_{\Lambda} \omega_{\mu}(\k) \nonumber\\[2mm]
&&\;\;= \frac{1}{\Omega} \int_{\rm B.Z.} \!\!d^2 k \; \sum_{\mu}
        \partial_{\Lambda} \; \omega_{\mu}(\k)\,, \label{Lambdacond}
\eea
where $\Omega=8\pi^2/\sqrt{3}$ is the volume of the unit cell. 
One finds that $Q$ and $\Lambda$ enter the 
Eq.~(\ref{Lambdacond}) only via the ratio $\xi=\Lambda/Q$. 

The requirement that the frequencies must be positive,  
$\omega_{\mu}(\k)\geq 0$, defines a lower limit $\xm$ for $\xi$: 
the frequencies $\omega_{\mu}(\k)$ are positive for $\xi > \xm$; for  
$\xi = \xm$, the lowest mode $\omega_{\mu_{0}}$ vanishes at some point(s) 
$\k_0$ in the BZ. 
When this happens, the corresponding condensate density 
${x}_{\mu_0}(\k_0)$ can be put non-zero, if this is necessary to satisfy 
Eq.~(\ref{Lambdacond}). 
It is important to note that in order to determine the actual value of 
$\xm$ (as well as those of $Q$, $\alpha$ and $\beta$)
it suffices to only consider Eq.~(\ref{Lambdacond}) at 
$x_{\mu}(\k)=0$, irrespective of whether there is condensate, 
$\omega_{\mu_0}(\k_0) = 0$, or not, $\omega_{\mu}(\k) \neq 0$ for all 
$\k,\mu$.

We solve the Eq.~(\ref{Lambdacond}) for $\xi$ numerically in two steps. 
First, we determine $\xm$: 
we decrease $\xi$ from large positive values until the condition 
$\omega_{\mu_0}(\k_0)=0$ signals that  $\xi = \xm$. 
Second, we set ${x}_{\mu}(\k)\equiv0$ and attempt to satisfy 
Eq.~(\ref{Lambdacond}) in the interval $\xi\geq\xm$. 
To this end, we set $\Lambda = \xi Q$ in Eq.~(\ref{EMFK})
and vary $\xi$ to determine the extremum of $\EMF$ 
(\ie Eq.~(\ref{Lambdacond})). 
We find that the extremum is a maximum.
If this maximum occurs for some $\xi>\xm$, then Eq.~(\ref{Lambdacond}) 
is satisfied with ${x}_{\mu}(\k)=0$.
If, however, $\EMF(\alpha,\beta, \xi Q, Q)$ decreases monotonously as
we lower $\xi$ down to $\xi=\xm$ , then the Eq.~(\ref{Lambdacond}) 
cannot be solved with ${x}_{\mu}(\k)=0$.
In this case, a finite condensate density ${x}_{\mu_0}(\k_0)\neq 0$, 
is required, in order to ``compensate'' for too large a value of the 
lhs.~of Eq.~(\ref{Lambdacond}). 
This fixes both $\xi=\xm$ and the value ${x}_{\mu_0}(\k_0)$ 
(cf. sects III B and IV B of Ref.~\onlinecite{S92}).

\paragraph{}
Having determined the value of $\xi$, we notice that the function 
$\EMF(\alpha,\beta,\Lambda,Q)$ is quadratic in $Q$ and bounded from below, 
which allows an analytical determination of $Q$ as the position of the 
minimum. 

\paragraph{} 
Finally, knowing the values of $\Lambda$ and $Q$, we proceed by a 
numerical extremalization of $\EMF$ with respect to the angles. 
The calculations show that $\EMF$ as a function of the angle $\beta$ 
possesses a maximum, and a minimum as a function of the angle $\alpha$
after $\beta$ has been eliminated. 
Thus, the variational energy $\EMF$ is bounded from below in 
the variables $Q_1$ and $Q_2$, as expected.

\paragraph{} 
We iterate this procedure ({\it i})-({\it iii}) until convergence is 
achieved.

\subsection{Numerical results of the Sp(N) formalism}
\label{results}
The results of the $\SPN$ approach in the large-$\CN$ limit are 
summarised in the zero temperature phase diagram of the AKAF, 
Fig.~\re{phasediag}. 
\begin{figure}[b]
\psfrag{J/(J+1)}{$J/(J+1)$}
\psfrag{1/kappa}[r][r]{$1/\kappa$}
\psfrag{IC/SRO}{{IC/SRO}}
\psfrag{IC/LRO}{{IC/LRO}}
\psfrag{DC/SRO}{{DC/SRO}}
\psfrag{DC/LRO}{\parbox{1cm}{DC/\\LRO}}
\psfrag{FM}{FM}
\includegraphics[width=8cm]{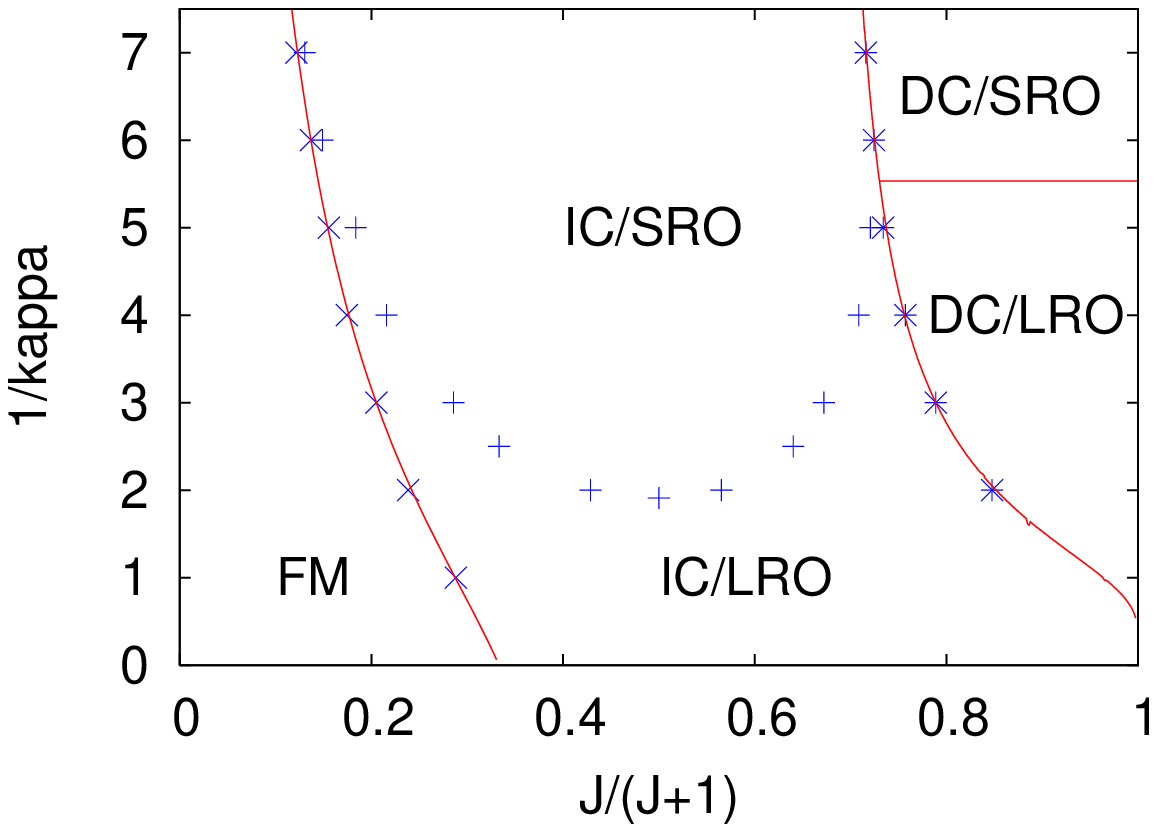}
\caption{\Co Phase diagram of $\HAKAF$ as obtained in the $\SPN$ approach.
Symbols and lines, respectively, denote numerical and analytical results 
for the phase boundaries (see text, Subsec.~\re{minimax} and 
Appendix \re{dc}).
Quantum fluctuations increase along the vertical axis. 
LRO: Long Range Order; SRO: short range order; FM: ferrimagnet; 
IC: incommensurate phase; DC: decoupled chains. 
At $J=1$, the results of Ref.~\onlinecite{S92} are recovered.
Incommensurate order (see Fig.~\re{qor}) occurs between the boundaries
of the ferrimagnetic phase ($\times$) and of the decoupled chain phase
($\ast$).}
\label{phasediag}
\end{figure}
The central part of the phase diagram is occupied by the incommensurate 
(IC) phase with LRO at sufficiently small $1/\kappa$.
The phase boundary that separates the region with SRO from the region 
with LRO was found by checking whether for a given pair of $J$ and 
$1/\kappa$ the lowest branch of the one spinon spectrum $\omega_{\mu}(\k)$ 
has zeros in the BZ or not, \ie whether there will be condensate at one 
or several points in the Brillouin zone or not. 
As one might expect, LRO is maximally suppressed by quantum fluctuations 
for $J=1$, which is the case of maximal frustration.

For $J=0$, the exact quantum ground state of the AKAF is ferrimagnetic 
(FM) according to the Lieb-Mattis theorem \cite{LM62}.
In this state, the expectation value $Q_1$ which measures the singlet 
weight on the horizontal bonds vanishes.
As shown in Fig.~\re{Q1}, our $\SPN$ calculations recover this exact 
result and extend it to a finite interval 
$0 \leq J \leq J_{ferri}(\kappa)$, which narrows as $1/\kappa$ increases.
\begin{figure}[t]
\psfrag{J/(1+J)}{$J/(J+1)$}
\psfrag{Q1}[r][r]{$Q_1$}
\includegraphics[width=8cm]{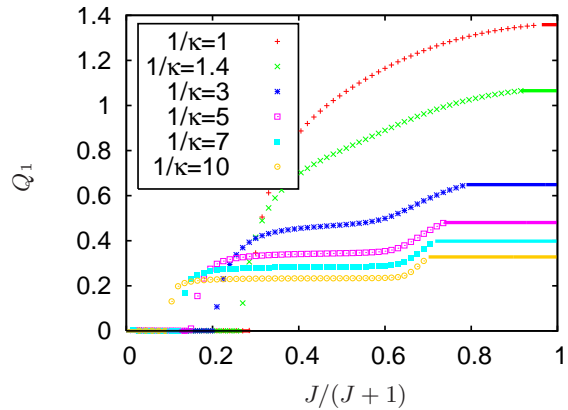} 
\caption{\Co Mean field parameter $Q_1$ as function of the anisotropy.} 
\label{Q1}
\end{figure}
\begin{figure}[b]
\psfrag{J/(1+J)}{$ J/(J+1)$}
\psfrag{Q2}[r][r]{$Q_2$}
\includegraphics[width=8cm]{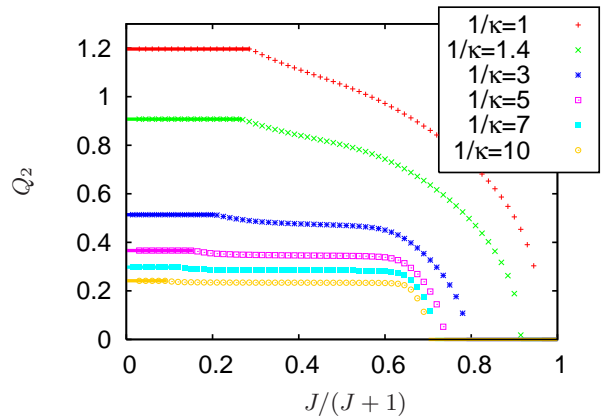}
\caption{\Co Mean field parameter $Q_2$ as function of the anisotropy.} 
\label{Q2}
\end{figure}
The parameter $Q_2$, which measures the singlet weight on the diagonal 
bonds, is independent of $J$ in this interval; its value decreases 
as $1/\kappa$ increases (see Fig.~\re{Q2}). 
Remarkably, the FM state retains its LRO in its entire region of 
existence.

As $J$ is increased beyond $J_{ferri}(\kappa)$, $Q_1$ increases in the 
manner of an order parameter at a second order phase transition. 
At the same time, the parameter $Q_2$ begins to decrease, and eventuallly 
it drops to zero at some $J = J_{DC}(\kappa)$.
Thus, the  large-$\CN$ approach predicts the existence of a 
decoupled-chain phase in the region above the phase boundary 
$J_{DC}(\kappa)$. 
$Q_2$  decreases to zero continuously so that the phase transition at 
$J_{DC}(\kappa)$ appears to be of second order again. 

Both LRO and SRO phases may be characterised by an ordering wave vector 
$\qord = 2\kmin$, where $\kmin$ is that wave vector at which the 
one-spinon excitation spectrum $\omega_{\mu}(\k)$ has its minimum. 
The static spin structure factor $S(\mathbf q)$  develops a peak at 
$\qord$.
\begin{figure}[t]
\psfrag{J/(1+J)}{$J/(J+1)$}
\psfrag{s1durchpi}{$q_{ord}^x/\pi$}
\includegraphics[width=8cm]{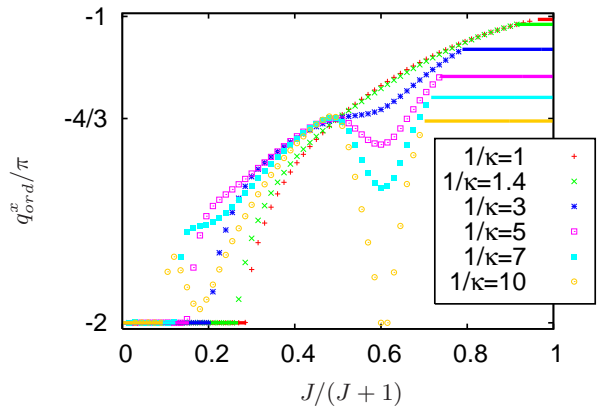}
\caption{\Co Ordering wave vector $q_{ord}^x$ as function of the anisotropy.} 
\label{qor}
\end{figure}
In Fig.~\re{qor}, we display the $x$-component of the ordering vector 
$q_{ord}^x = q_{ord}^x(J)$ ($q_{ord}^y=0$).  
At the {\kago} point $J=1$,  $|q_{ord}^x| = 4\pi/3$ is independent  
of the value of $\kappa$.
For $1/\kappa \lesssim 3$, the behaviour of $q_{ord}^x$ as a function of 
$J$ is as expected: as $J$ increases, it increases monotonously until 
the phase boundary $J_{DC}(\kappa)$ is reached and remains constant inside 
the DC phase. 
However, for $1/\kappa \gtrsim 3$ the function $q^{ord}_ x(J)$ develops 
a minimum at $J \approx 1.5$, which becomes more pronounced as $1/\kappa$ 
increases. 

In Sec.~\re{minimax} we emphasised that contrary to previous 
applications of the large-$\CN$ approach to spin models on {\kago} and
anisotropic triangular lattices \cite{SR91,S92,CMK01},
we found it essential to consider two chemical potentials $\la$ and $\lc$ 
here, one for the spins on the horizontal lattice lines ($\la$) and one 
for the middle spins ($\lc$). 
We display the values of these parameters as functions of $J$ in 
Fig.~\re{lambda}. 
\begin{figure}[b]
\psfrag{J/(1+J)}{$J/(J+1)$}
\psfrag{lambda}[r][r]{$\la$, $\lc$}
\psfrag{la}[r][r]{$\la$}
\psfrag{lc}[r][r]{$\lc$}
\includegraphics[width=8cm]{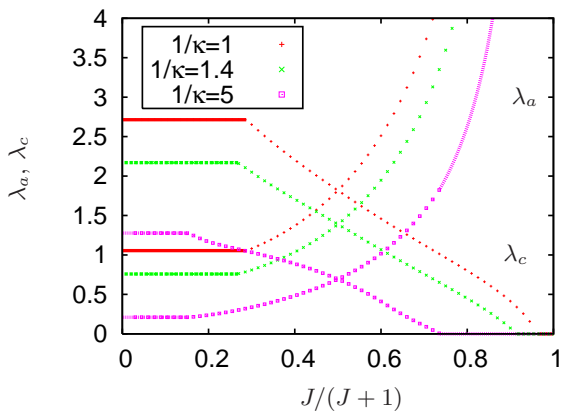}
\caption{\Co Lagrange multipliers $\la$, $\lc$ (chemical potentials)
as functions of the anisotropy.} 
\label{lambda}
\end{figure}
We have no physical explanation for the behaviour of $\la$, $\lc$  
as functions of $J$ and $\kappa$ but it is gratifying to see that  
$\la = \lc$ at  $J = 1$ independent of $\kappa$ in accordance with  
earlier work \cite{S92}.

As indicated above, along with numerical study of Eq.~\re{EMF}, 
we performed extensive analytical calculations, both to corroborate the
numerics and to obtain additional insight into the problem.
Details of the analytical techniques are presented in Appendix \ref{dc}. 
Here we state that we were able to analytically determine $\SPN$ 
phase boundaries between the SRO and LRO DC phase, between the DC and 
IC phase, and between the FM and IC phase, see Fig.~\re{phasediag}.
Moreover, our analytical calculations allowed us to explicitly confirm the 
existence of LRO inside the FM phase and immediately to the right of 
the FM-IC phase boundary.
Likewise, the regions with SRO and LRO inside and immediately to the left 
of IC-DC phase boundary were determined analytically.
This was achieved by evaluating in these regions the $\SPN$ generalisation 
of the spin-spin correlation function  
$\langle \S_{i,u} \cdot \S_{j,v} \rangle$ of the model defined by 
expression (\re{SPNCORR}). 
($u, v = a, b, c$ denote the sites of the triangular cells $i$ and $j$ 
of the model, see Fig.~\re{QP}). 
On the right-hand side of the FM-IC boundary and inside the FM phase, 
we find for large distances between the cells, 
$|\mathbf{r}_j   -  \mathbf{r}_i| \gg 1$,

\begin{subequations}
\be
\langle \S_{i,u} \cdot \S_{j,v} \rangle \sim S_u \; S_v \;,
\label{CORFMA}
\ee
where
\bea
S_w \sim \sqrt{ \frac{3}{2} }
\frac{|x_3 (\kmin)|^2}{\ND\,(\lambda_c + \lambda_a)} \;
\left\{\begin{array}{ll} \frac{\lambda_c}{2} & \quad w = a, b\\[2ex]   
                          -\lambda_a & \quad w = c  \end{array}   \right.
\label{CORFMB}
\eea
\label{CORFM}
\end{subequations}

and $u,v = a, b, c$ denote the sites of the triangular cells $i$ and $j$
of the model, see Fig.~\re{QP}. 
Here, $|x_3 (\kmin)|^2/\ND$ is the condensate density at 
$\kmin = (- \pi,0)$,\quad $|x_3 (\kmin)|^2/\ND = \kappa$, 
see Eq.~(\re{DELAM-}). 
On the FM-IC transition line and inside the FM phase, where 
Eqs.~(\re{CORFM}) are valid, the parameters $\la$ and $\lc$ 
are not independent but can  be expressed in terms of a single parameter 
$\delta$, see Eqs.~(\re{QUOTDEL}), (\re{OM3=0}).
The sign pattern on the right-hand side of  Eq.~(\re{CORFM}) and the 
ordering wave vector $\mathbf{q}_{ord} = 2 \kmin = (-2\pi,0)$ are indeed 
the properties one expects to find for the long-distance behaviour of the 
spin-spin correlation function of a ferrimagnetically ordered state. 
Since $|x_3 (\kmin)|^2/\ND$ remains finite for arbitrarily small values of 
$\kappa$, the mean-field $\SPN$ approach predicts that this order persists 
in the extreme quantum limit of our model, $1/\kappa \gg 1$. 
Together with Eqs.~(\re{CORFM}), the fact that 
the condensate density $|x_3 (\kmin)|^2/\ND$ remains 
constant inside the FM region, see Eq.~(\re{DELAM-}),  
implies that the magnetisation of the FM phase remains 
constant up to the FM-IC phase boundary. 
The same behaviour of the magnetisation of a ferrimagnetic phase 
has previously been observed in an exact-diagonalisation study of 
a one-dimensional {\kago}-like antiferromagnet \cite{WKSME00}. 
At the FM-IC phase boundary, the magnetisation becomes spatially modulated 
with an incommensurate wave vector $\qord = 2\kmin$.\\    
On the left-hand side of the IC-DC boundary and inside the DC phase 
we find the following large distance behaviour 
of the spin-spin correlation function:

\begin{subequations}
\bea
\langle \S_{i,c} \cdot \S_{j,c} \rangle\, &\sim& \,\frac{3}{2} 
\cos \left[2 \,\kmin\, ( \mathbf{r}_i   -  \mathbf{r}_j) \right]
\left(\frac{2q_1^2}{1+q_1^2} \right)^2\nonumber \\[2mm]
&&
\cdot \; \left[ \frac{ |x_3 (\kmin)|^2 + 
|x_3 (-\kmin)|^2 }{\ND\,q_2^2 \lambda_a \omega^{(2)}_3(\kmin)}\right]^2\,, 
\label{CORDCA}\\[4mm]
\langle \S_{i,u} \cdot \S_{j,v} \rangle\, &\sim&\, 0\qquad \mbox{for} \quad 
u,v\, \ne\, c,c\,.
\label{CORDCB}
\eea
\label{CORDC}
\end{subequations}
Here, $q_1$ and $\la$ denote the saddle-point values of these variables 
obtained from Eqs.~(\re{q1}), (\re{lama}). 
$q_2$ is a function of $q_1$, determined by Eq.~(\re{Q2Q1}) or by 
Eq.~(\re{q1q2}) depending on whether $1/\kappa < 1/\kappa_s$ or  
$1/\kappa > 1/\kappa_s$
($\kappa_s=0.181$, see Fig.~\re{IandK}).
$\omega^{(2)}_3(\kmin)$ is the value of the second-order expansion 
coefficient of the lowest spinon frequency $\omega_3(\k)$, 
cf. Eqs.~(\re{OMEXP2}), (\re{OMEXP32}), at its minimum, 
and $2\,\kmin$ is the ordering wave vector immediately to the left on the 
IC-DC phase boundary and inside the DC phase; 
it is determined by Eq.~(\re{KMIN}).  
$|x_3 ( \kmin)|^2/\ND = |x_3 ( -\kmin)|^2/\ND$  are the condensate 
densities at the wave vectors $\pm \kmin$. 
As is shown in Appendix \ref{dc},  $\omega^{(2)}_3(\kmin)$ remains finite 
for $1/\kappa > 1/\kappa_s$ and hence $|x_3 (\pm \kmin)|^2/\ND$ vanishes. 
Thus, $\langle \S_{i,c} \cdot \S_{j,c} \rangle \sim 0$, 
\ie there is no LRO in this region. 
By contrast, for $1/\kappa < 1/\kappa_s$ both, $|x_3 ( \pm \kmin)|^2/\ND$  
and $\bar{\omega}^{(2)}_3(\kmin)$ vanish when the IC-DC phase boundary 
is approached from the left. 
However their ratio, which determines the spin-spin-correlation function,
Eqs.~(\re{CORDC}), remains finite in this limit according to 
Eq.~(\re{X3}). 
Thus, for $1/\kappa < 1/\kappa_s$, Eqs.~(\re{CORDC}) show that while 
the chain spins $\S_a$, $\S_b$ remain disordered,  
there is long-range IC order between the middle spins $\S_c$  along the
IC-DC phase boundary and inside the DC phase for sufficiently large $\kappa$. 
The middle spins occupy the sites of a triangular lattice.
Remarkably, the correlations between these spins predicted by 
Eqs.~(\re{CORDC}) are compatible with the spin pattern 
\be
\S_{j,c} = S\left[\cos(2\kmin \mathbf{r}_j) \hat{\mathbf{e}}_x 
+ \sin(2\kmin \mathbf{r}_j)\hat{\mathbf{e}}_y \right]
\ee
that would obtain if the middle spins $\S_c$ were classical spins coupled 
by a classical Heisenberg model with exchange constant 
$\tilde J$ along one lattice direction and couplings $\tilde{J'}$ along 
the other two directions with a ratio $\tilde{J'}/\tilde J$ such 
that incommensurate order with wave vector $2\kmin$ would be established.
 This persistence of long range order in the DC phase of the AKAF distinguishes our result 
from the result obtained by  Chung {\it et al.}, Ref.~\onlinecite{CMK01}, 
in  their large-$\CN$ 
$\SPN$  treatment of the anisotropic triangular antiferromagnet:  there the DC phase 
consists of uncorrelated linear spin chains. Qualitative considerations of the 
finite-$\CN$ corrections 
to the mean-field $\SPN$ result led the authors of Ref.~\onlinecite{CMK01}
 to the conclusion that 
instead of the DC phase there is spin-Peierls order in the large--$J$ 
region of their model. In the next section, we will present a different approach, a block-spin 
perturbation theory, to get further insight into the properties of 
the AKAF for the physical spin-1/2 case.

\section{Block-spin perturbation approach}
\label{quantum}
\begin{figure}
\includegraphics[width=6cm]{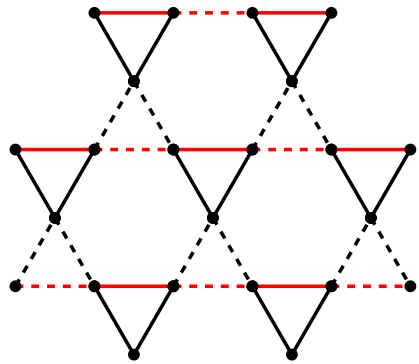}
\caption{\Co The {\kago} lattice as a triangular lattice of downward 
pointing triangles.
The coupling strength is $J$ on the horizontal bond and unity on the other 
two bonds.}
\label{trimkag}
\end{figure}

The basic idea of the block-spin perturbation theory is to calculate 
the states of small clusters of a given lattice exactly and to treat 
the coupling between these clusters perturbatively. 
The basic building blocks of the {\kago} lattice are triangles. 
Thus it is natural to consider the trimerised {\kago} lattice in which 
the spins on the downward pointing triangles are assumed to be strongly 
coupled whereas the coupling on the bonds of the upward pointing triangles 
are assumed to be weak, see Fig.~\ref{trimkag}. 
(Clearly, the exchange of the roles of the upward and the downward 
pointing triangles will not affect the further development to be 
presented in the current section.)
The Hamiltonian for this trimerised model reads

\be
\CH(J, \gamma) = \CH_{\bigtriangledown}(J) 
               + \gamma \CHt(J)\,,\,\,\, 0 \leq \gamma \leq 1\,,
\label{GAKAGO}
\ee
where $\CH_{\bigtriangledown}(J)$ \;($\CHt(J)$) denote those terms in 
the Hamiltonian (Eq.~\re{akago}) that act on the bonds of the 
downward (upward) pointing triangles.
We will determine approximate GSs of this trimerised model in 
different ranges of $J$ in a perturbation expansion w.r.t.~$\gamma$. 
The hope is  that the results will provide some qualitative insight 
into the GS properties of the non-trimerised model $\CH(J, 1)$ which 
is our original model Eq.~(\re{akago}). 
The same strategy has previously been applied sucessfully to frustrated 
spin models by several authors \cite{S95,M98,RRRS00,Z05}.\\

Obviously, the GSs of the unperturbed Hamiltonian $\CH(J,0)$ are 
products of GSs of the individual downward pointing triangular 
plaquettes. 
The GSs of a single plaquette and the corresponding energies are\\ 

\def\theparagraph{\arabic{paragraph}}
\setcounter{paragraph}{0}
\paragraph{for $J < 1$:}
\begin{subequations} 
\be
\ket{\alpha} =  \frac{1}{\sqrt{6}} \Big[ 
\big(\ket{\uparrow \uparrow \downarrow} - \ket{\downarrow \uparrow \uparrow}\big) + 
\big(\ket{\uparrow \downarrow \uparrow} - \ket{\downarrow \uparrow \uparrow}\big)
\Big]\,,
\label{alpha}
\ee
\be
\ket{\bar{\alpha}} =  \frac{1}{\sqrt{6}} \Big[ 
\big(\ket{\downarrow \downarrow \uparrow } - \ket{\uparrow\downarrow \downarrow }\big) + 
\big(\ket{\downarrow \uparrow \downarrow} - \ket{\uparrow \downarrow \downarrow} \big)
\Big]\,,
\label{baralpha}   
\ee
\be
\varepsilon_{\alpha} = \varepsilon_{\bar{\alpha}} =  -1 + J/4 
 \label{Ealpha}\,;
\ee 
\label{alp}
\end{subequations}

\paragraph{$J > 1$:}
\begin{subequations} 
\be
\ket{\beta} = \frac{1}{\sqrt{2}} 
\big(\ket{\uparrow \downarrow \uparrow} - \ket{\uparrow \uparrow \downarrow} \big)\,,
\label{beta}
\ee
\be
\ket{\bar{\beta}} = \frac{1}{\sqrt{2}} 
\big(\ket{ \downarrow \uparrow \downarrow} - \ket{\downarrow \downarrow \uparrow} \big)\,,
\label{barbeta}
\ee
\be
\varepsilon_{\beta} = \varepsilon_{\bar{\beta}} = -3/4 J\,.  \label{Ebeta}
\ee
\label{bet}
\end{subequations}

\begin{figure}
\psfrag{P1}[r][r]{\large $\ket{\alpha (\bar{\alpha})}$:}
\psfrag{P2}[r][r]{\large $\ket{\beta(\bar{\beta})}$:}
\psfrag{sp}[cb][l]{$\uparrow(\downarrow)$}
\includegraphics[width=6cm]{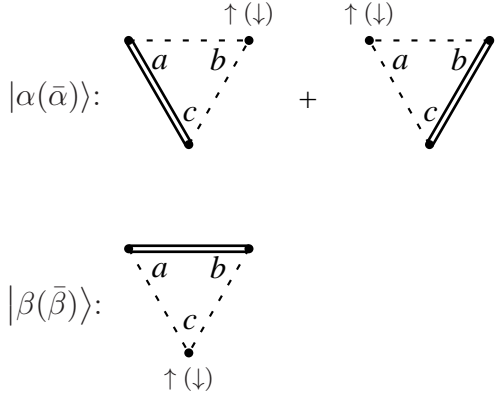}
\caption{Ground-states of triangular plaquettes. 
Heavy lines depict singlets. 
The coupling strength is $J$ on the horizontal bond and unity 
on the other two bonds.}
\label{plaq}
\end{figure}

Here, the ket vectors denote the spin state of the plaquette 
in the $S^z$ basis. 
The three arrows inside the $\ket{cba}$ symbol denote from left to right 
the spin direction at the sites $c$, $b$ and $a$ of the plaquettes in 
Fig.~\re{plaq}. 
The states $\ket{\alpha}$\; ($\ket{\bar{\alpha}}$) and $\ket{\beta}$\; 
($\ket{\bar{\beta}}$) have total z-spin $1/2$\; ($-1/2$). 
They can be depicted graphically as shown in Fig.~\re{plaq}. 
From these plaquette states, the zeroth order GSs of the Hamiltonian
$\CH(J, \gamma)$ will be constructed. 
We treat the cases $J < 1$ and $J > 1$ separately.\\   

\setcounter{paragraph}{0}
\paragraph{$J < 1$:}
Since the states $\ket{\alpha}$, $\ket{\bar{\alpha}}$ are the 
GSs of the individual downward pointing plaquettes in this case, the states

\be 
\ket{A(M)} = \prod_{i \epsilon \{M\}} \ket{\alpha_i}  
                \prod_{j \epsilon \{\ND-M\}} \ket{\bar{\alpha}_j}\,,            
\label{A}
\ee

are here the zeroth order GSs of $\CH(J, \gamma)$.
The set $\{M\}$ is a subset of $M$ out of the $\ND$ downward pointing 
triangles of the $3\ND$-site {\kago} lattice; the subscripts $i$, $j$ 
denote the position of individual triangular plaquettes
in the lattice of these plaquettes which is also triangular, 
see Fig.~\re{trimkag}. 
The zeroth order energy eigenvalues associated with the states 
$\ket{A(M)}$ do not depend on $M$:

\be
E^{(0)}_{A(M)} = \ND(-1 + J/4)\,.
\label{EAM}
\ee

Hence, there are in total $2^{\ND}$ degenerate zeroth order GSs 
$\ket{A(M)}$.
The single plaquette states $\ket{\alpha}$, $\ket{\bar{\alpha}}$ satisfy 
the conditions for the validity of the Lieb-Mattis theorem, 
Ref.~\onlinecite{LM62}: after a canonical transformation which rotates the 
spins on the sites $a$ and $b$ by $\pi$ around the z-axis  
$\ket{\uparrow} \rightarrow i \ket{\uparrow}$,  
$\ket{\downarrow} \rightarrow -i \ket{\downarrow}$, and which leaves 
the spins on the site $c$ fixed the coefficients of all basis states 
on the right sides of Eqs.~(\re{alpha},\,\re{baralpha}) become positive 
($+1/\sqrt{6}$). 
As a consequence,  all the GSs $\ket{A(M)}$ satisfy the conditions 
for the validity of the Lieb-Mattis theorem. 
For $J=0$ it follows from this theorem that the total magnetisation 
of the {\em exact} quantum GS $\ket{\Phi_{exact}}$ of the Hamiltonian 
$\HAKAF$ must be an eigenstate of the total magnetisation
\be
\hat{m}_{tot} = 
\sum_i^{\ND} (S^z_{i,a} + S^z_{i,b} + S^z_{i,c} )
\label{mag}
\ee 
with eigenvalue $m_{tot} = \ND/2$, \ie $\ket{\Phi_{exact}}$ must be 
a ferrimagnetic state. 
By continuity, one expects this to be the case not only for $J=0$, 
but up to a certain finite value of $J$.
This suggests that the state $\ket{A(M\!=\!0)}$, c.f.~Eq~(\re{A}), is the 
appropriate {\em zeroth} order GS in this case and that the 
degeneracy of the states $\ket{A(M)}$ is lifted by the perturbation 
$\CH_{\bigtriangleup}$ in favour of the state $\ket{A(0)}$. 
To confirm this, we determine the creation energy of a flipped plaquette 
in first order in $\gamma$, \ie the difference of the energy of the 
state with one plaquette spin flipped relative to the ferrimagnetic state, 
and the energy of the ferrimagnetic state:
\be
\delta^{(1)} E_A({M\!=\!1)} = E_{A(1)} -  E_{A(0)}. 
\label{EDELA1}
\ee
A simple calculation yields 
\be
\delta^{(1)}E_A({M\!=\!1)} = \frac{4}{9} \gamma (1 - J)\,,     
\label{RESEDELA1}
\ee
\ie to first order, $\ket{A(M\!=\!0)}$, the ferrimagnetic GS 
is stable w.r.t.~a flip of a single plaquette spin, as long as $J < 1$.

As a further check on the stability of the state $\ket{A(M\!=\!0)}$, we  
calculate the dispersion of the excitation  energy of a propagating single 
flipped plaquette spin. 
For this purpose, we need to determine the overlap matrix elements between 
the state with a flipped plaquette spin at the site $j$ and states 
with a flipped spin at one of the neighbouring sites, 

\begin{subequations} 
\be 
t_{j,j\pm\fdel_1} = \bra{\bar\alpha_j} \bra{\alpha_{j\pm\fdel_1}} 
  \gamma J \S_j \S_{j\pm\fdel_1} 
  \ket{\alpha_j} \ket{\bar\alpha_{j\pm\fdel_1}} 
=  \frac{2}{9} \gamma J\,,\\
\ee
\be
t_{j,j\pm\fdel_{2,3}} = \bra{\bar\alpha_j} \bra{\alpha_{j\pm\fdel_{2,3}}} 
\gamma \S_j \S_{j\pm\fdel_{2,3}} 
  \ket{\alpha_j} \ket{\bar\alpha_{j\pm\delta_{2,3}}} 
= -\frac{1}{9} \gamma\,.
\ee 
\end{subequations}
Here, $\del_{\nu},\;\nu\,=\,1,\,2,\,3,$ are the primitive lattice vectors 
of the {\kago} net, see Fig.~\re{aniskago}; 
they connect the sites of the plaquette lattice.
Then, by diagonalising the ensuing transfer Hamiltonian 
{\setlength{\arraycolsep}{0pt}
\bea 
\CH_{trans} = \gamma \sum_j 
  \Big\{ &&\frac{2}{9} J \left(\, \ket{j + \del_1} \bra {j }
                     + \ket{j - \del_1} \bra {j }\, \right) \NL{0mm}
&& -  \frac{1}{9}  \left( \ket{j + \del_2} \bra{j} + 
        \ket{j - \del_2} \bra{j} \right) \NL{2mm}
&& -  \frac{1}{9} \left( \ket{j + \del_3} \bra{j}  
        + \ket{j - \del_3} \bra{j} \right) \,  \Big\}\,,
\label{htrans}
\eea}

where $\ket{j}$ denotes the state with a flipped plaquette spin at site 
$j$, we obtain for the kinetic energy of this excitation: 
\be
\varepsilon({\k}) = \frac{4}{9} \gamma 
\left[J \cos(k_x) - \cos(\frac{k_x}{2}) \cos(\frac{\sqrt{3} k_y}{2}) \right]\,.
\ee
Adding the  energy for the creation of a single flipped plaquette spin, 
we find for the total energy of the excitation in the limit of small 
wave vector $\k$ 
\be
\omega({\k}) = \frac{2}{9} \gamma  
  \left[\, (\frac{1}{4} - J) k^2_x + \frac{3}{4} k^2_y 
  + {\mathcal O}(k^4)\, \right] \;.
\label{omega}
\ee
 
Obviously, the ferrimagnetic state $\ket{A(M\!=\!0)}$ becomes unstable 
against a {\em propagating} flipped plaquette spin already at $J = 1/4$, 
\ie much earlier than suggested  by the excitation energy of a 
{\em static} flipped spin (see Eq.~(\re{RESEDELA1})).
We remark that this bound is independent of the actual magnitude of the 
perturbation parameter $\gamma$ and therefore, the qualitative result may 
survive in the limit $\gamma =1$.\\
  
\paragraph{$J > 1$:}

In this region, the states 

\be
\ket{B(M)} = \prod_{i \epsilon \{M\}} \ket{\beta_i}
                \prod_{j \epsilon \{\ND-M \}} \ket{\bar{\beta}_j}
\label{B}
\ee

with eigenenergy 

\be
E^{(0)}_{B(M)} = \ND(-3J/4)\,.
\label{EBM}
\ee
 
are the zeroth order eigenstates of $\CH(J, \gamma)$.
These states consist of free spins on the $c$-sites and of spin-singlet 
dimers that cover every second bond of the horizontal chains of the 
lattice. 
We wish to answer the question of whether the $2^{\ND}$-fold degeneracy of 
these states, which results from the degrees of freedom of the free spins, 
is lifted by the perturbation $\gamma \CH_{\bigtriangleup}$; 
in other words, we want to find out whether the middle spins remain 
decoupled from the chain spins.
We proceed as in case ({\it i}). We compare in a perturbation expansion 
w.r.t.~$\gamma$ the energy of the state $\ket{B(0)}$ with the energy of 
$\ket{B(1)}$, {\it i.e.} with the state with one plaquette spin flipped 
relative to $\ket{B(0)}$. 
We denote this difference by 
$\delta^{(1)} E_B(M\!=\!1) = E_{B(1)} - E_{B(0)}$.
Surprisingly, we find that the matrix elements 
$\bra{B(M)}\CHt\ket{B(M)}$ vanish for any choice of $M$. 
There is no first order correction to the energy $E^{(0)}_{B(M)}$, 
$\delta^{(1)} E_B(M\!=\!1) = 0$.
Moreover, we observe that the off-diagonal matrix elements 
$\bra{B'(M)}\CHt\ket{B(M)}$, where $\ket{B'(M)}$ and 
$\ket{B(M)}$ contain identical numbers of states $\ket{\beta}$, 
$\ket{\bar{\beta}}$ but differ in their distribution over the $\ND$ 
downward pointing triangles, also vanish. 
This implies that, in contrast to case ({\it i}), a flipped plaquette spin 
cannot hop to a neighbouring site in a first order process.
Coupling between the spins on the $c$-sites  occurs only in second order 
in $\gamma$. 
It is succinctly described by an effective spin Hamiltonian for the 
$c$-site spins, which are at the same time total spins of the downward 
pointing plaquettes (see Fig.~\ref{plaq}):

{\setlength{\arraycolsep}{0cm}
\bea
\CH_{eff}= \sum_{i\,\in\, c}\;\sum_{\nu\,=\,1}^3 && \Big\{  
  J_{\fdel_{\nu}}^{\|} \;S_i^z\,S_{i+\fdel_{\nu}}^z \NL{0mm}
&+& J_{\fdel_{\nu}}^{\bot} 
  \left(S_i^x\,S_{i+\fdel_{\nu}}^x\,+\,S_i^y\,S_{i+\fdel_{\nu}}^y \right) 
\Big\}\,. 
\label{HCEFF}
\eea}\\[-2ex]
Here, $S_i^{\alpha}$, $\alpha = x$, $y$, $z$, denote plaquette 
spin operators; $i$ is the position of a downward pointing plaquette 
on the triangular lattice formed by these plaquettes. 
The exchange couplings $J_{\fdel_{\nu}}^{\|}$ and $J_{\fdel_{\nu}}^{\bot}$
are given as second order matrix elements of $\CH_{\bigtriangleup}$: 

\begin{subequations}
\bea 
J^{\|}_{\fdel_{\nu}}&=&\gamma^2 
 \left\{ \sum_X \frac{ 
   \bra{B_{i \uparrow,\, i' \uparrow }} \CHt \ket{X} \, 
   \bra{X} \CHt \ket{B_{i \uparrow,\, i' \uparrow }}}
  {2\,\varepsilon_B - \varepsilon_X} \right. \NL{0mm}
 &&\left. - \sum_Y \frac{
   \bra{B_{i \uparrow,\, i' \downarrow}} \CHt \ket{Y} \, 
   \bra{Y} \CHt \ket{B_{i \uparrow,\, i' \downarrow}}}
  {2\,\varepsilon_B - \varepsilon_Y} \right\} 
\label{JPAR}
\eea
\bea
J^{\bot}_{\fdel_{\nu}}&=&\gamma^2  
\sum_X \frac{
   \bra{B_{i\downarrow,\,i'\uparrow}} \CHt \ket{X} \, 
   \bra{X} \CHt \ket{B_{i\uparrow,\, i' \downarrow}}}
  {2\,\varepsilon_B - \varepsilon_X}\,,
\label{JPERP}
\eea
\label{JALL}
\end{subequations}

and $i'\!\equiv\!i\!+\!\del_{\nu}$. 
Here, the states $\ket{B_{i\sigma,\, i' \sigma'}}$ are zeroth order 
GSs, Eq.~(\re{B}), whose spin patterns are identical on all sites 
except for the sites $i$ and $i'$ where the $z$-components of the spins 
take the values $\sigma$ and $\sigma'$, respectively; 
$\ket{X}$ and  $\ket{Y}$ are excited states of $\CH_{\bigtriangledown}$. 
Of course, since the $\SUZ$ symmetry of the original Hamiltonian 
$\CH(J, \gamma)$ must be conserved in the derivation of $\CH_{eff}$, 
the expressions Eqs.~(\re{JALL}) must yield identical results, 
$J_{\fdel_{\nu}}^{\|}\,=\,J_{\fdel_{\nu}}^{\bot} \equiv J_{\fdel_{\nu}}$.
Non-zero contributions to $J^{\|}_{\fdel_{\nu}}$ and 
$J^{\bot}_{\fdel_{\nu}}$ are obtained if either the same term  
$\S_i \S_{i'}$ of $\CHt$ acts in both matrix elements of the numerators 
of Eq.~(\re{JALL}) (two-block contributions) or the terms 
$\S_i \S_k$,\; $\S_k \S_{i'}$ act in the left and right elements, 
respectively, where the plaquette geometry must be as shown in 
Fig.~\re{heff} (three-block contributions). 
In contrast to the case of the isotropic KAF studied by Zhitomirsky 
\cite{Z05}, the three-block contributions do not produce three-spin 
interactions in the present case. 
Rather, they contribute to the exchange interactions $J_{\fdel_1}^{\|}$ 
and  $J_{\fdel_1}^{\bot}$ of the Hamiltonian $\CH_{eff}$, Eq.~(\re{HCEFF}). 
The evaluation of the expressions (\re{JALL}) yields
\bea
J_{\fdel_1} &=&\;\;  \gamma^2\, \frac{1}{288}\,\frac{1}{J} 
\left[ \frac{56}{1 - \frac{1}{J}} 
- \frac{1}{1 - \frac{1}{4J}} + \frac{98}{1 + \frac{1}{2J}} \right] \NL{3mm}
&=&\quad \frac{17}{32}\frac{\gamma^2}{J}\,
 \left[ 1 + \mathcal{O}(J^{-1}) \right]
\label{JEFF1} 
\eea
and
\bea
J_{\fdel_2}\,=\,J_{\fdel_3}\,&=&\, 
\frac{\gamma^2}{6J}\left[\frac{1}{1 - \frac{1}{J}} \, - \, 
\frac{1}{1 + \frac{1}{2J}} \right] \nonumber\\[3mm]
&=&\,\frac{\gamma^2}{4J^2} \left[1\,+\,\mathcal{O}(J^{-1}) \right]\,. 
\label{JEFF2}
\eea

\begin{figure}[t]
\psfrag{i}[bc][bl]{$i$}
\psfrag{is}[bc][bl]{$i'$}
\psfrag{k}[bc][bl]{$k$}
\includegraphics[width=7cm]{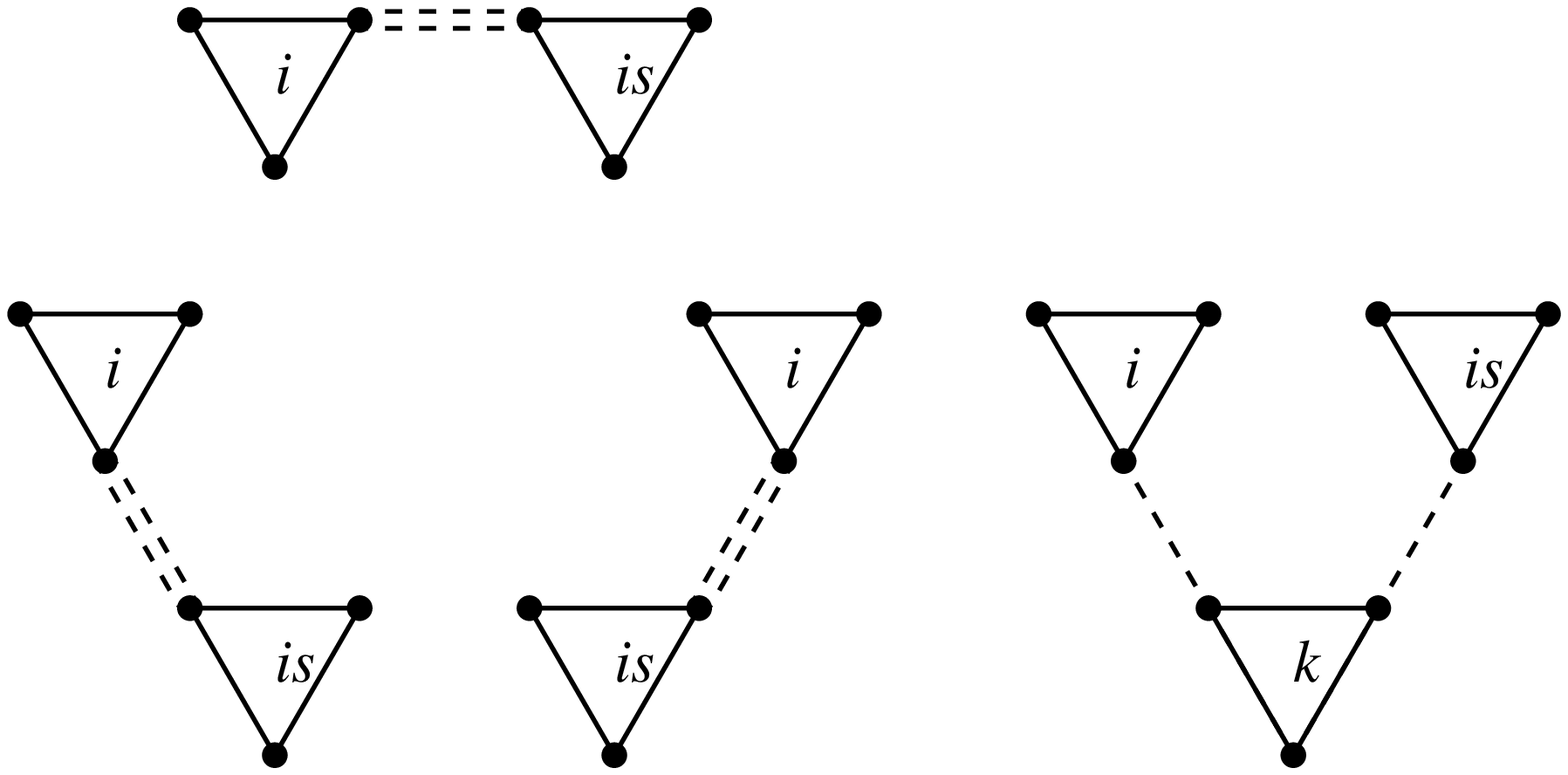}
\caption{Configurations of $\bigtriangledown$ blocks contributing to the 
interblock couplings $J_{\delta_\nu}^{\|}$ and $J_{\delta_{\nu}}^{\bot}$. 
Double dashed lines indicate that the same term element of 
$\CH_{\bigtriangleup}$ acts twice between the $\bigtriangledown$ blocks 
at sites $i$ and $i'$ (see also text).} 
\label{heff}
\end{figure}

\begin{figure}[b]
\includegraphics[width=0.2\textwidth]{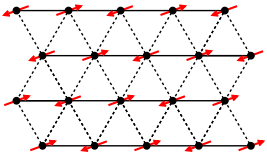}
\caption{\Co Collinear antiferromagnetic state (CAF) on the triangular lattice
\cite{SB07}.}  
\label{caf-drei}  
\end{figure}

\begin{figure}[t]
\includegraphics[width=6cm]{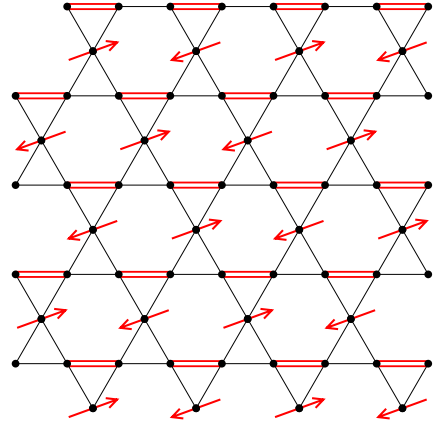}
\caption{\Co Tentative ground-state of the anisotropic {\kago} 
antiferromagnet in the limit $J \gg 1$. 
Double lines: dimers between the spins on the end points.}
\label{trim_gs}
\end{figure} 

Obviously, these results are useful for $J\,\gg 1$. 
There, $\CH_{eff}$ represents a spin $1/2$ Heisenberg Hamiltonian on the 
triangular lattice of the $c$-sites with a coupling along the $\del_1$ 
direction that is strong ($\mathcal{O}(\gamma^2/J)$) in comparison to the 
couplings in the two other directions ($\mathcal{O}(\gamma^2/J^2)$).  
This limiting case of the anisotropic triangular Heisenberg antiferromagnet (ATHAF)
has recently been analysed by Starykh and Balents with field theoretical 
methods \cite{SB07}. 
These authors find that in the limit of strong anisotropy, 
$K \equiv J_{\fdel_1}/J_{\fdel_2} \rightarrow \infty$, 
the GS of the model Eq.~(\re{HCEFF}) is a {\em collinearly} ordered   
antiferromagnet (CAF) as depicted in Fig.~\ref{caf-drei}. 
Since the ordering wave vector $\q_{CAF}\,=\, (\pi, \pi/2)$ of this phase 
does not evolve continuously from the ordering wave vector 
$\q_{IC}$ of the incommensurate (IC) {\em spiral} phase of the ATHAF, 
($\q_{IC}\,=\,(q_x(K), 0)$ with  $-3\pi/2 \leq q_x(K) \leq -\pi$ for $1/2 \leq K \leq \infty$),  
they conclude that the IC phase and the CAF phase must be separated 
by a quantum phase transition. 
For the trimerised anisotropic {\kago} model, Eq.~\re{GAKAGO},  
these results have the following implications:\\
{\it i)} While  in the limit of strong 
anisotropy $J \gg 1$ there is long-range 
collinear antiferromagnetic order among the $c$-site spins, 
 the $a$- and $b$-site spins are paired in singlets, see Fig.~\ref{trim_gs}.\\ 
{\it ii}) This picture of the GS of the trimerised anisotropic {\kago} model Eq.~(\re{GAKAGO}) 
differs from the result obtained in the $\SPN$ approach insofar as for sufficiently large 
$\kappa$ the $\SPN$ approach predicts 
long range IC order among the $c$-site spins  up to 
arbitrarily large values of $J$.  Thus, if the picture of a CAF phase for large anisotropy 
 persists in the non-trimerised limit $\CH(J,\gamma=1)$ of the model Eq.~\ref{GAKAGO}, 
one would expect a quantum phase transition between the IC phase and the CAF phase 
of the AKAF similarly as for the ATHAF. 
In closing this section, we remark that the calculation that led to the effective  
Hamiltonian $\CH_{eff}$, Eq.~\re{HCEFF}, {\ie} to the coupling between the $c$ 
spins in the strongly anisotropic limit, shows clearly that this coupling arises 
from quantum fluctuations of the $a$ and $b$ spins.

\section{Summary and discussion}
\label{summary}
In this work, we have studied the ground state (GS) phase diagram of the 
quantum Heisenberg antiferromagnet on the {\kago} lattice with spatially 
anisotropic exchange (AKAF). 
The model is relevant for a description of magnetic properties of 
volborthite, which is a natural realisation of a spin $1/2$ antiferromagnet 
consisting of weakly coupled slightly distorted {\kago} layers.
A small monoclinic distortion along one of the three lattice directions 
causes the exchange coupling along this direction, $J$, to differ 
from the couplings in the other two directions, $J'$,
which we set equal to unity, cf. Fig.~\ref{aniskago}.
We have investigated the problem in the full range of the anisotropy, 
$0 \leq J \leq \infty$,  using three different approximate methods: 
the classical and semiclassical approach, 
the mean-field $\SPN$ approach, and a block-spin perturbation theory.

The case $J = 1$ is the much studied isotropic {\kago} antiferromagnet 
(KAF). 
Exact diagonalisation studies of this model \cite{LBLPS97, WEBLSLP98} are 
available. 
Their results speak conclusively in favour of a spin liquid ground state 
\cite{ML04}.
This view is supported by block-spin approaches \cite{M98, MM00}. 
Conflicting results have been found in 
Refs.~\onlinecite{MZ91, ZE95, SM02, NS03, BA04},
where various valence bond crystal (VBC) states are proposed 
as ground states of the KAF. 
However, a recent comparison of the exact spectrum of the 36-site sample 
of the KAF against the excitation spectra allowed by the symmetries 
of these states, casts doubts on their validity \cite{MS06}.   

Within the whole anisotropy range, the case $J=0$ is special, since it 
allows for an {\it exact} characterisation of the quantum GS as 
ferrimagnetic (FM) with a total magnetisation of $M = S\,N_s/3$ for a 
system of $N_s$ spins of magnitude $S$.
In the classical picture, this state corresponds to a unique staggered 
layout of spins with a non-zero net magnetisation of the lattice unit cell 
(cf.~Fig.~\re{FerriKago}).
In the classical limit, the ferrimagnetic ground state survives up 
to $J= 1/2$. 
For $J>1/2$, the ``chain'' spins (\ie spins coupled by $J$) begin to tilt 
gradually towards the middle (remaining) spins (see Fig.~\re{cansp}).
This allows for a formation of a large degenerate manifold of canted spin 
states. 
In contrast to the isotropic case $J=1$, where the degeneracy 
grows exponentially with the system size $N_s$, its growth is weaker:
$2^{1.26\sqrt{N_s}}$ for $J\not=1$. This implies that there must 
be an increasingly large number of classical low energy configurations 
as $J$ approaches unity.
In the linear semiclassical approximation, the spin-wave spectrum 
has one zero-frequency  mode for each point of the magnetic Brillouin zone 
(BZ).
The spectrum is identical for the different canted states for all $J>1/2$. 
Thus, in this order of the semiclassical approximation, no 
order-by-disorder mechanism appears that would select one particular state
or a particular group of states from the manifold of canted states 
as true ground states. 
In the limit $J \rightarrow \infty$, the frequency spectrum of 
non-zero modes gradually takes the shape of the spectrum that one would 
expect for a set of uncoupled antiferromagnetic spin chains parallel to 
the strong-$J$ direction. 
No qualitative change from the set of canted spin states to the set 
of decoupled chains at a finite value of $J$ is found.

We have further explored the nature of the phases at various $J$ 
exploiting the mean field (MF) $\SPN$ approach, that incorporates the 
effect of quantum fluctuations not only perturbatively, but 
self-consistently. 
The strength of quantum fluctuations is controlled by a parameter 
$\kappa$, which is the analogue of the spin value $S$ in the original 
$\SUZ$ symmetric model.
In fact, for $\CN = 1$, when the $\SP{1}$ symmetric model is equivalent to 
the $\SUZ$ model, $\kappa = 2S$. 
For general $\CN$, this last identity does not hold, but $\kappa$ is 
still a measure for the importance of quantum fluctuations that are 
strong for $\kappa \ll 1$  and weak for $\kappa \gg 1$.
In the MF $\SPN$ approach, the nature of the phases that occur can be read 
from the values of the mean field parameters $Q_1$ and $Q_2$ and 
from the spectrum of the bosonic spinon excitations. 
While the mean field parameters $Q_1$ and $Q_2$ (cf. Fig.~\ref{QP}) are	 
the GS expectation values of singlet bond operators,
the  structure of the spinon spectrum, $\omega_{\mu}(\k;Q, \lambda)$,
determines the existence or non existence of long-range order (LRO): 
If the spectrum becomes gapless at some wavevector $\qord$, 
a Bose condensate will form and a modulated structure with 
the wavevector $2\qord$ will acquire LRO.

As was to be expected, the phase diagram of the AKAF 
obtained by the MF $\SPN$ approach contains an incommensurate (IC) phase 
in the vicinity of the isotropic point $J=1$ which is ordered for 
sufficiently large $\kappa$ according to this approach, see 
Fig.~\re{phasediag}. 
Qualitatively, we may gauge the value of $\kappa$ against the spin 
length $S$ by looking at the line $J=1$ of the phase diagram which is the 
location of the $\SPN$ analogue of the isotropic $\SUZ$ symmetric 
{\kago} model: 
since, as we have argued above, the $\SUZ$ model is disordered for 
$S=1/2$, we may conclude from Fig.~\re{phasediag} that the value of 
$1/\kappa$ that corresponds to $S=1/2$ must be greater than two.
Somewhat surprisingly, the FM phase remains long-range ordered for 
arbitrarily small $\kappa$. 
This may reflect the fact that in the $\SUZ$ version of the model,
the FM phase is ordered even for the smallest physical spin value $S=1/2$.
A new feature of the phase diagram is the prediction  of a decoupled chain 
(DC) phase for large enough $J$, which has no classical analogue.
In this phase, the chains of strongly coupled $a$- and $b$-site spins 
show no magnetic order. 
The $c$-site spins which are interspersed between these chains and 
which occupy the 
sites of triangular sublattice are decoupled from the chain spins. However, they may or 
may not exhibit long range order among themselves depending on the magnitude of $\kappa$.
Remarkably, the spin-spin correlations, whose asymptotics were obtained analytically, 
are compatible with the spin-spin correlations of an anisotropic classical Heisenberg 
antiferromagnet
on the triangular lattice whose exchange couplings differ in one direction from those in
the other two directions.

In order to tackle the problem of the GSs of the AKAF
from a third corner, we have used a block-spin perturbation theory. 
This method has the advantage of being applicable directly to the spin 
$1/2$ version of the model. 
In applying this approach, one has to initially group the spins of 
the model in clusters. 
For the {\kago} lattice, it is natural to choose the spins around either 
the upward or the downward pointing triangles as clusters of strongly 
coupled units and to consider the coupling between these clusters, 
$\gamma$, as the small expansion parameter. 
Thus one trimerises the original model (see Fig.~\re{trimkag}) and in so 
doing, one breaks the translational invariance of the original model. 
In the zeroth order of this expansion, two regions can be distinguished 
by the eigenenergies of the individual trimers: $J<1$ and $J>1$. 
For sufficiently small $J$, one recovers the FM state as the GS 
in first order w.r.t.~$\gamma$. 
For $J>1$, there are no first order corrections to the energy. 
Following  an earlier application of the block-spin technique to the 
isotropic KAF \cite{Z05}, we determine for $J>1$  in second order  
in $\gamma$ an effective Hamiltonian $\CH_{eff}$ for the block-spins which 
can be identified as the middle spins of the original model and 
that occupy the sites of a triangular lattice. 
$\CH_{eff}$ is a Heisenberg Hamiltonian with a coupling $J_{\fdel_1}$ 
of the order of $\gamma^2/J$ along the $\del_1$ direction 
(cf. Fig.~\re{aniskago}) and couplings $J_{\fdel_2} =  J_{\fdel_3}$ 
of the order of $\gamma^2/J^2$ along the other two directions. 
The calculations that lead to these results show clearly that the 
couplings between the $c$-spins of the AKAF are due to fluctuations 
of the singlets between the $a$- and $b$-spins into excited states. 
In a very recent field theoretical study, Starykh and Balents \cite{SB07}
arrive at the conclusion that for $J_{\fdel_1} \gg J_{\fdel_2, \fdel_3}$, 
the ground state of the anisotropic triangular antiferromagnet  
represented by $\CH_{eff}$ is a collinearly ordered spin state, see Fig.~\re{caf-drei}.
Then, together with the singlet dimers between the $a$- and $b$-spins 
of the downward pointing triangles, the state depicted in  
Fig.~\re{trim_gs} emerges as the candidate ground state of the AKAF in the 
limit of large anisotropy $J \gg 1$: 
while nearest neighbour spins on the strongly coupled $a$-$b$ chains 
form singlets and decouple magnetically from the spins on the $c$ sites,
the latter order in a collinear antiferromagnetic structure.
This structure cannot be obtained by a continuous deformation of the spiral IC structure 
that is predicted by the $\SPN$ approach and is believed to prevail 
for sufficiently large $\kappa$ in the region of moderate anisotropy. 
As a consequence of the trimerisation, the state depicted in  
Fig.~\re{trim_gs} breaks the translational symmetry of our original model, 
Eq.~(\re{akago}). 
If this state survives as the ground state of the non-trimerised model, 
{\ie} when the expansion parameter $\gamma$ approaches unity,
then, owing to their incompatible symmetries, the spiral IC phase and 
the large J phase of our model must be separated by 
a quantum phase transition.

\begin{acknowledgments}
One of the authors (HUE) acknowledges a useful discussion with F.~Mila.
The work at the University of Waterloo was supported by the Canada 
Research Chair (Tier I, Michel Gingras).
We thank M. Gingras for a critical reading of the manuscript and numerous 
helpful suggestions. 
\end{acknowledgments}

\appendix

\section{Ground state degeneracy for general $J$ }
\label{gsd}
\begin{figure}[b]
\psfrag{c1}{\large $\chi_1$}
\psfrag{c2}{\large $\chi_2$}
\psfrag{c3}{\large $\chi_3$}
\psfrag{c4}{\large $\chi_4$}
\psfrag{c5}{\large $\chi_5$}
\psfrag{c6}{\large $\chi_6$}
\includegraphics[width=6cm]{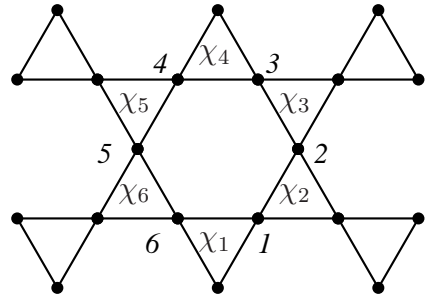}
\caption{Chiralities around heaxagonal plaquette}
\label{hexa}
\end{figure}

We first derive the constraint on the chiralities that leads to the 
reduction in the number of degenerate ground states for general $J$ 
relative to the special case $J = 1$. 
Let  $\chi_1,\, \cdots,\, \chi_6$ be the chiralities of the six triangles 
surrounding one of the hexagons of the {\kago} lattice, and let 
$\phi_1,\, \cdots, \, \phi_6$ denote the angles that define the directions 
of the spin vectors on the six corners of the hexagon, see Fig.~\ref{hexa}. 

Then, as is seen in Fig.~\ref{hexa}, the following relations between 
the angle $\phi_1,$ and the angles $\phi_2 \, \cdots, \,\phi_6$ are an 
immediate consequence of these definitions:
\begin{subequations} \jot2ex
\bea
\phi_2 &=& \phi_1 -\theta\,\chi_2\,,  \\
\phi_3 &=& \phi_1 -\theta\,(\chi_2 + \chi_3)\,, \\   
\phi_4 &=& \phi_1 -\theta\,(\chi_2 + \chi_3) -(2\pi \!-\! 2\theta)\;\chi_4\,,\\
\phi_5 &=& \phi_1 -\theta\,(\chi_2 + \chi_3 + \chi_5) -(2\pi \!-\! 2\theta)\;\chi_4\,,\\ 
\phi_6 &=& \phi_1 -\theta\,(\chi_2 + \chi_3 + \chi_5 + \chi_6) -
          (2\pi \!-\! 2\theta)\;\chi_4  ,\qquad \\
&&\mbox{and}\quad \phi_6 = \phi_1 +(2\pi - 2\theta)\chi_1\,.
\eea
\end{subequations}

\begin{figure}[t]
\includegraphics[width=7cm]{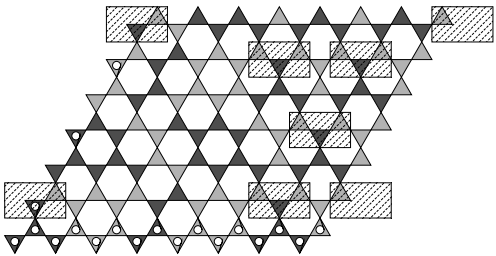}
\caption{Example of a chirality distribution; dark and light shaded 
triangles represent positive and negative chirality, respectively. 
Chirality configurations in boxes fix the chirality distribution of the 
row above them uniquely. 
An empty circle inside a triangle indicates that its chirality can be 
chosen freely to be positive or negative.}
\label{chideco}
\end{figure}

From the last two of these relations it follows that the chiralities 
$\chi_1,\, \cdots\,, \chi_6$ are constrained by the sum rule 
\be
\chi_2 + \chi_3 + \chi_5 + \chi_6 - 2\,\chi_1 - 2\,\chi_4 = 0\,.
\label{CONST}
\ee  
For the isotropic {\kago} system, $J = 1$, $\theta = 2\pi/3$, 
one finds instead of the constraint (\re{CONST}) the sum rule 
\be 
\sum_{j=1}^6 \chi_j = n \quad \mbox{where} \quad n = 0,\, 1,\, 2
\label{CONSTISO}
\ee
which is obviously less restrictive than (\re{CONST}).\\
Next, we present the arguments that lead to the estimate 
\be
N^{aniso}_{GS}(\ND) \lesssim 2^{\alpha \sqrt{\ND}} \quad \mbox{with} 
 \quad \alpha < 3
\label{NUMDISTRI}
\ee
for the number $N^{aniso}_{GS}(\ND)$ of classical GSs of an anisotropic 
{\kago} AF with $\ND$ downward pointing triangles (the number of sites 
is $3 \ND$). 
Any planar configuration of a cell of the {\kago} lattice can be constructed 
by decorating the successive rows of up and down pointing triangles 
with chirality values $\chi = \pm 1$ starting with the first row.
We consider only square cells with $\sqrt{\ND}$ rows with $\sqrt{\ND}$ 
downward pointing triangles. 
Then, each row consists of $2\sqrt{\ND}$ triangles, 
see Fig.~\ref{chideco}. 

\begin{figure}[t] 
\psfrag{log Anzahl Zust.}{$\ln N^{aniso}_{GS}$}
\psfrag{sqrt(Ns/3)}{$\sqrt{\ND}$}
\includegraphics[width=7cm]{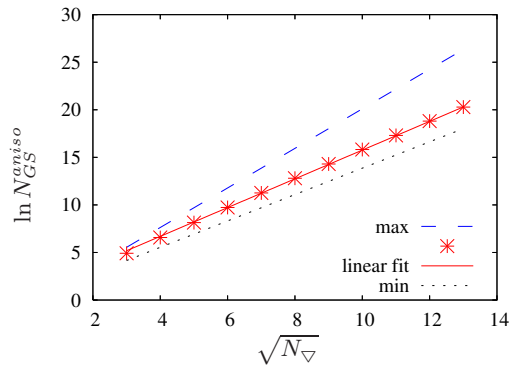}
\caption{\Co Number of chirality distributions, $N_{GS}^{aniso}$,  
of cells of up to $\ND=13 \times 13$. 
Dotted line: $\mbox{min} = 2\,\sqrt{\ND}\,\ln2$ (lower bound); 
dashed  line: $\mbox{max} = 3\, \sqrt{\ND}\,\ln2$ (upper bound);
full line: $\ln(N_{GS}^{aniso}) = 0.65 + 2.18 \,\sqrt{\ND}\,\ln2$ 
(linear fit to the numerical results).}
\label{numdistri}
\end{figure}

Obviously, there are $2^{2\sqrt{\ND}}$  ways to decorate the first row. 
Disregarding certain exceptions, which will be discussed below, one can, 
for a given configuration of the first row, choose the chirality of an 
arbitrary triangle of the second row to be either $+1$ or $-1$.
After this choice has been made, the constraint (\re{CONST}) fixes the 
chiralities of all the remaining triangles of the second row uniquely. 
Proceeding in this manner from row to row one would generate 
$2^{2\sqrt{\ND}}\,\cdot\,2^{\sqrt{\ND}}$ distributions of chiralities 
over the $\ND$ downward pointing triangles of the cell. 
For finite lattice cells, the requirement of periodic boundary conditions 
imposes further constraints on the number of possible chirality distributions 
in these cells, but the effect of these constraints will become negligible 
in the thermodynamic limit $\ND \rightarrow \infty$. 
However, there is a further reduction of the number of possible chirality 
distributions:    
For a given distribution in a row it is not {\em always} possible to find 
{\em two} distributions for the successive row which both satisfy the 
constraint (\re {CONST}). 
If in a row the lower half of a hexagon of the next row is decorated 
by chiralities in the manner $-\, +\, -$ or  $+\, -\, +$ (see boxes in 
Fig.~\ref{chideco}), then the chiralities of the next row are fixed uniquely. 
This reduces the number of possible chirality distributions. 
Obviously, this reduction of the number of possible chirality distributions 
survives in the thermodynamic limit so that the exponent in (\re{NUMDISTRI}) 
is less than $3\,\sqrt{\ND}$, the value one would have expected without 
this reduction. 
We have calculated the number of distributions for cells of up to 
$\ND=13 \times 13$ and have found the value $\alpha \simeq 2.18$ 
for the constant in the expression (\re{NUMDISTRI}), see 
Fig.~\ref{numdistri}.

As we have mentioned above, the sum rule (\re{CONSTISO}) which applies for 
the isotropic {\kago} AF is less restrictive than the sum rule (\re{CONST}). 
Consequently, the number of chirality distributions in the isotropic 
model \cite{B70},
\be
N_{GS}^{iso} \sim 1.1833^{3\ND}
\label{NUMISO}
\ee
is larger than in the anisotropic model. 
Since the transition from the anisotropic model to the isotropic model happens 
through a continuous variation of the coupling constant $J$, there should be 
a continuous transition between the numbers of GS configurations in these two 
cases. 
Presumably, this transition implies that the density of low-energy states 
of the anisotropic model increases exponentially with an exponent 
$\sim \sqrt{\ND}$ so that for $J \rightarrow 1$ a sufficient number of states 
collapses to the GS to bring about the transition between the laws 
(\re{NUMDISTRI}) and (\re{NUMISO}).

\section{Phase boundaries}
\label{dc}
The FM phase and the DC phase are  chacterised by the vanishing of the 
parameters $Q_1$ and $Q_2$, respectively.
Our numerical results in section \ref{results} show that at the respective 
phase boundaries, $Q_1$ and $Q_2$ decrease to zero like order parameters 
at second order phase transitions. 
This suggests that we expand the mean field energy $\EMF$, 
Eq.~(\ref{EMFK}), w.r.t.~either $Q_1$ or $Q_2$ in the manner of a 
Landau-Ginzburg (LG) expansion and determine the phase boundaries and the 
properties of the FM and the DC phase  from this expansion. 
We write $E_{MF}/(\ND \CN) = e^{(\alpha)}_{LG}(Q_{\alpha})$
where
\be 
e^{(\alpha)}_{LG}(Q_{\alpha}) = e_{\alpha} + r_{\alpha}\,|Q_{\alpha}|^2 + 
g_{\alpha}\,|Q_{\alpha}|^4 + \mathcal{O}(|(Q_{\alpha}|^6)\,.
\label{ELG}
\ee
The coefficients $e_{\alpha}$, $ r_{\alpha}$ and $g_{\alpha}$ are 
functions of the variables $\kappa$ and $J$, of the parameters $\la$, 
$\lc$ and of $Q_2$, $Q_1$ for $\alpha = 1,\,2$, respectively. 
The saddle point of $e^{(\alpha)}_{LG}(Q_{\alpha})$ w.~r.~t.~ $\la$, 
$\lc$ and $Q_{\beta}$, $\beta \not= \alpha$, determines the physical 
values of these parameters. 
For $e^{(\alpha)}_{LG}(Q_{\alpha})$ to qualify as a {\it bona fide} 
Landau-Ginzburg energy describing a second order phase transition with 
$Q_{\alpha}$ playing the role of an order parameter, the coefficients 
$g_{\alpha}$ have to be positive at the saddle point. 
For $g_{1}$, \ie inside and on the boundary of the FM phase, this follows 
from the numerical result: $Q_1$ is found to remain zero for all 
$J \leq J_{F}(\kappa)$. 
By contrast, we have no numerical results for $J\geq J_{DC}(\kappa)$, 
\ie inside and on the boundary of the DC phase. 
Therefore, we need to show by analytic means that $g_{2} > 0$. 

\subsection{The FM phase and the FM-IC phase boundary}

Since, as we have just remarked, we know that $g_{1}>0$, the remaining task is 
to determine the coefficients $e_1$ and $r_1$ of  $e^{(1)}_{LG}$. 
To this end, we have to expand the mean field energy $\EMF$, Eq.~(\ref{EMFK}), 
w.r.t.~$Q_1$ which amounts to expanding the frequencies  
$\omega_{\mu}(\k)$ w.r.t.~$Q_1$. 
As can be inferred from the expressions (\ref{DP}), (\ref{MP}) the frequencies 
depend on 
$Q_1$ only through the combination $\varepsilon^2 = J^2 |\tilde{Q}_1|^2$. 
Therefore, we write the expansion in the form
\bea
\omega_{\mu}(\k;\varepsilon)\,\, &=&\,\, \omega_{\mu}^{(0)}(\k) + 
\varepsilon^2 \omega_{\mu}^{(1)}(\k) + \mathcal{O}(\varepsilon^4) \NL{0mm}
	 &=& \,\, \lambda_+ \left[\bar{\omega}_{\mu}^{(0)}(\k) + 
	\bar{\varepsilon}^2 \bar{\omega}_{\mu}^{(1)}(\k)  + 
\mathcal{O}(\bar{\varepsilon}^4) \right]
\label{OMEXP}
\eea
with $\lambda_+ = ( \lambda_a +  \lambda_c)/2\,,\quad
\bar{\omega}_{\mu}^{(0)} =  \omega_{\mu}^{(0)}/\lambda_+\,, \quad 
\bar{\omega}_{\mu}^{(1)}(\k) = 
\lambda_+\,\partial_{\varepsilon^2} \,\omega_{\mu}(\k;\varepsilon) 
  |_{\varepsilon = 0}$ 
\quad and \quad $ \bar{\varepsilon} = \varepsilon/\lambda_+$.

Here, the introduction of the ``dimensionless'' quantities 
$\bar{\omega}_{\mu}^{(i)}$, $\bar{\varepsilon}$ looks like an unneccessary 
complication but it will help to keep expressions further below simple.
Setting $Q_1 = 0$ in the matrix $\hat{\mathbf{D}}(\omega)$, Eq.~(\ref{DP}),  
and solving Eq.~(\ref{DETD}) for $\omega$ we find 
\begin{subequations}
\bea
\bar{\omega}_1^{(0)}(\k)\,\, &=& \,\, \wf + \delta \,,\\
\label{OMEG1}
\bar{\omega}_2^{(0)}(\k)\,\, &=& \,\, 1 - \delta \,,\\
\label{OMEG2}
\bar{\omega}_3^{(0)}(\k)\,\, &=& \,\, \wf - \delta \,.
\label{OMEG3}
\eea
\end{subequations}
Here
\be
 \delta\, = \,\lambda_-/\lambda_+\quad \mbox{with}\quad 
  \lambda_- = (\lc - \la)/2
\label{QUOTDEL}
\ee
and 
\be
\wf\, = \,\sqrt{1 - \bar{q_2}^2[\sin^2(s_2/2) + \sin^2(s_3/2)]}
\label{wf}
\ee
with
\bea
&&\bar{q}_2 = |Q_2|/\lambda_+\,\\
&&\mbox{and}\quad\,\, s_a \,= \,{\del}_a \mathbf{k}\,,\,\, a = 2,\,3 \quad
  (\mbox{see Fig.~\ref{QP}})\, .  \nonumber
\label{BARQ2}
\eea

From our numerical results, Fig.~\ref{lambda}, we know that $\lc > \la$ and 
hence $\delta > 0$. 
Therefore, $\bar{\omega}_3^{(0)}(\k) < \bar{\omega}_{1,2}^{(0)}(\k)$,
and hence, if the minimum of $\bar{\omega}_3^{(0)}(\k)$ vanishes at the  
point $\kmin$  in the Brillouin zone,
$\bar{\omega}_{1,2}^{(0)}(\kmin)$ will be finite. 
Thus, since condensate can only occur when one of the frequencies 
$\bar{\omega}_{\mu}^{(0)}\,,\; \mu = 1,\,2,\,3$ vanishes there may be a 
finite condensate density  $|x_{3}(\kmin)|^2$, but the densities  
$|x_1|^2$ and $|x_2|^2$ will certainly be zero. 
With these remarks and with the above results for $\bar{\omega}_{\mu}^{(0)}$ 
we find from Eq.~(\ref{EMFK})

\bea
 e_{1}/\lambda_+ \;&=&\; 2 \lambda_+ \bar{q}_2^2 - (3 - \delta) (\kappa + 1) 
 \NL{0mm} 
 &&\; + \frac{1}{\ND} \sum_{\k} \left[ \bar{\omega}_{1}^{(0)}(\k) +  
 \bar{\omega}_{2}^{(0)}(\k) + \bar{\omega}_{3}^{(0)}(\k) \right] \NL{0mm}
 &&\; + \bar{\omega}_{3}^{(0)}(\kmin)\, |x_{3}(\kmin)|^2 /\ND\,.
\label{E01}  
\eea

Stationarity of $e_{1}$ w.r.t.~$ \lambda_-$, $ \lambda_+$, and 
$\bar{q}_2^2$ (which is equivalent to stationarity w.r.t.~$ \la$, $ \lc$), and 
$Q_2^2$ requires the following three conditions to be fulfilled:
{\setlength{\arraycolsep}{0mm}
\bea
&& \partial e_{1}/\partial \lambda_-  
= 0: \NL{2mm}
&&\mbox{\hspace{7mm}} \frac{1}{\ND}|x_{3}(\k_{min})|^2 = \kappa\,; 
\label{DELAM-} \\[5mm]
&&\partial e_{1}/ \partial \lambda_+ 
= 0: \NL{2mm}
&&\mbox{\hspace{7mm}} 
2\,\lambda_+\bar{q_2}^2 -\frac{3}{2} \kappa -1 + {\mathbf E}_2(\bar{q_2}) + 
  \frac{\kappa}{2} w_F(\kmin) = 0\,; \NL{0mm}
\label{DELAM+} \\[3mm]
&& \partial e_{1} /\partial \bar{q}_2^2 = 0: \NL{2mm}
&&\mbox{\hspace{7mm}}
2\,\lambda_+ - \frac{1}{\bar{q_2}^2} 
 \left[{\mathbf K}_2(\bar{q_2}) - {\mathbf E}_2(\bar{q_2}) \right]
- \frac{\kappa}{w_F(\kmin)} = 0\,; \NL{0mm}
\label{DEQ2F} \\[5mm]
&&\mbox{\hspace{7mm}}\mbox{with}\quad 
{\mathbf K}_2(\bar{q_2}) = \frac{1}{\pi}\int_0^{\pi}\!\!ds_2\,\,
 \frac{1}{\pi} \int_0^{\pi}\!\!ds_3 \; \wf^{-1}  \,, \NL{2mm}
&&\mbox{\hspace{7mm}}\mbox{\hspace{10mm}}
{\mathbf E}_2(\bar{q_2}) = \frac{1}{\pi}\int_0^{\pi}\!\!ds_2\,\,
 \frac{1}{\pi} \int_0^{\pi}\!\!ds_3 \; \wf  \,. \NL{0mm}
\eea}

According to Eq.~(\re{DELAM-}), condensate must be present in the FM region. 
This requires that  $\bar{\omega}_3^{(0)}(\kmin)$ vanishes. 
From Eq.~(\re{OMEG3}) it is seen that 
$\kmin = (-\pi,\,0)$, so that $\bar{\omega}_3^{(0)}(\kmin) = 0$, if
\be
 w_F(\kmin) \;=\; \sqrt{1-2\bar{q_2}^2} \;=\; \frac{\lambda_-}{\lambda_+}\,.
\label{OM3=0}
\ee
Within the FM region and on the FM-IC boundary (\ie for $Q_1 = 0$) 
the saddle-point values of $\bar{q_2}$, $\lambda_+$, and $\lambda_-$ 
are then determined as functions of $\kappa$ by the Eqs.~(\re{DELAM+}), 
(\re{DEQ2F}) and (\re{OM3=0}). 
Remarkably, within this region these quantities are independent of 
the value of the exchange constant $J$. 
The solution of these equations shows that $0 \leq \bar{q_2} \leq 2/3$ for 
$0  < \kappa < \infty$, cf.~Figs.~\re{Q2}, \re{lambda}.

The FM-IC phase boundary is the solution of $r_1(\kappa, J) = 0$ 
(cf.~Eq.~(\re{ELG}), where 

\be
\left. r_1 = \partial e^{(1)}_{LG} /\partial Q_1^2 \right|_{Q_1 = 0}
\label{R1}
\ee

with $e^{(1)}_{LG}$ ($\EMF$) from  Eq.~(\ref{EMFK}). 

We obtain
\bea
\frac{r_1}{J^2}\; &=&\; \frac{1}{J} 
 - \frac{1}{\lambda_+}\,\frac{1}{\ND}\,\sum_{\k}
 \sin^2\left(\frac{s_2 + s_3}{2} \right) \Omega^{(1)}(\k)\NL{0cm} 
&&  + \frac{\kappa}{\lambda_+}
   \lim_{\k \rightarrow \k_{min}} \left(\sin^2\left(\frac{s_2 + s_3}{2} \right)
   \bar{\omega}_{3}^{(1)}(\k) \right) \label{R1J} 
\eea

with\qquad $\Omega^{(1)}(\k) 
  = - \bar{\omega}_{1}^{(1)}(\k) - \bar{\omega}_{2}^{(1)}(\k) 
    - \bar{\omega}_{3}^{(1)}(\k)$. 

To obtain the expansion coefficients $\bar{\omega}_{\mu}^{(1)}(\k)$ which 
appear in the last equation, we solve Eq.~(\re{DETD}) to first order 
in the expansion w.r.t.~$\bar{\varepsilon}^2$. We find 
\bea
\Omega^{(1)}(\k) 
  &=& \left. \frac{1}{\wf(\wf + 1 -2\delta)} \right[ \wf + 1  \NL{2mm}
  &{}& \left. + \frac{\bar{q_2}^4\,\sin^2\left(\frac{s_2}{2}\right) 
		\,\sin^2\left(\frac{s_3 }{2}\right)\,
		(2 \wf + 1 - \delta)}{(\wf + 1)
		(1 - \delta)\,(\wf^2 - \delta^2)} \right]  \NL{0mm}
\label{OM1}
\eea             
and 
\be
\lim_{\k \rightarrow \kmin}\left( \sin^2  \left(\frac{s_2 + s_3}{2} \right)\,
 \bar{\omega}_{3}^{(1)}(\k)\right)
= -\frac{1}{2\delta} \,\frac{1 + \delta}{1 - \delta}\, .
\label{DOM1}
\ee

With these results Eq.~(\re{R1J}) can, in the thermodynamic limit, 
be cast into the form
\be
\frac{r_1}{J^2}  = \frac{1}{J}\, -\, \frac{I_3(\bar{q_2})}{\lambda_+} 
  - \frac{\kappa}{\lambda_+}\,  \frac{1}{2\delta}\, 
   \frac{1 + \delta}{1 - \delta}  
   = \frac{1}{J} - \frac{1}{J_{F}(\kappa)}\,,   \label{JQ2}
\ee
where
\be
I_3(\bar{q_2}) = \frac{1}{\pi}\!\int_0^{\pi}\!\!\!ds_2\, 
                 \frac{1}{\pi}\!\int_0^{\pi}\!\!\!ds_3\, 
                	  2 \sin^2\left( \frac{s_2}{2} \right) 
			    \cos^2\left( \frac{s_3}{2} \right) 
			   \Omega^{(1)}(\k) \, .
\ee

Then, with $\bar{q}_2 = \bar{q}_2(\kappa)$ and 
$\lambda_{\pm} = \lambda_{\pm}(\kappa)$ 
as obtained from Eqs.~(\re{DELAM+}), (\re{DEQ2F}) and (\re{OM3=0}), 
the condition $r_1\,=\,0$ is an equation for the FM-IC phase boundary 
$J=J_{F}(\kappa)$ which yields the graph shown in Fig.~\re{phasediag}.   
As we have mentioned above, inside the FM region, \ie for 
$J < J_{F}(\kappa)$, the saddle-point values of the quantities $\bar{q}$ 
and $\lambda_{\pm}$ and hence of $Q_2$, $\la$, $\lc$ and $|x_3(\kmin)|$ 
are independent of the exchange coupling $J$, \ie they retain the values 
they attain on the FM-IC phase boundary, cf.~Figs.~\re{Q2}, \re{lambda}. 

\subsection{The DC phase and the IC-DC phase boundary}

Proceeding in exact analogy to the development in the previous subsection 
we now expand $\EMF/(\ND \, \CN)$ in powers of $|Q_2|^2$. However, instead 
of working with the variables $Q_1$, $Q_2$, $\lambda_a$, $\lambda_c$ we work 
with $q_1$, $Q_2$, $\lambda_a$, $q_2$ here, where
 
\begin{subequations}
\bea
q_1 &=& \frac{J |Q_1|}{\la}\,,\label{defq1}\\
q_2 &=& \frac{ |Q_2|}{\sqrt{\la \lc}}\,\label{defq2}.
\eea
\end{subequations}

The replacement of $|Q_1|$ is purely a matter of convenience. 
By contrast, the replacement of variables  $Q_2$, $\lc$, which according to 
the numerics vanish simultaneously as $J$ approaches the IC-DC phase boundary, 
by the pair $Q_2$, $q_2$ leaves us with only one vanishing variable, since, 
as will be seen below, $q_2$ remains finite throughout. 

\subsubsection{Expansion of $e^{(2)}_{LG}(Q_2)$}
We write

\be 
\omega_{\mu}(\k)\, =\, \omega^{(0)}_{\mu}(\k)\, + \, \omega^{(2)}_{\mu}(\k)\, Q_2^2\, + 
\,\omega^{(4)}_{\mu}(\k)\,  Q_2^4\, + \mathcal{O}( Q_2^6)
\label{OMEXP2}
\ee      

and determine the coefficients $\omega^{(n)}_{\mu}$, $n\, = \,1,\,\dots\, 4,$ 
by solving Eq.~(\ref{DETD}) for $\omega$ iteratively. We obtain
\begin{subequations}
{\setlength{\arraycolsep}{-2mm}
\bea
&&\omega^{(0)}_{1}(\k) + \omega^{(0)}_{2}(\k)\, =  2\la \, \wdc \,, \quad 
\omega^{(0)}_{3} = 0 \,,
\label{OMEXP20} \\[3mm]
&&\omega^{(2)}_{1}(\k) + \omega^{(2)}_{2}(\k)\,= \,-\frac{1}{\la}\, 
\frac{1-\cos k^x\,\cos k^y}{\wdc} \,, \label{OMEXP22}\\[3mm]
&&\omega^{(2)}_{3}(\k)\, = \frac{1}{q_2^2\,\la}\,\left[ 
C(\k)^2 - D(\k)^2 \right]^{1/2} \!\!. \label{OMEXP32}
\eea} \label{OMEXPALL}
\end{subequations}
Here,
\be
 C(\k)\;=\; 1\,-\,q_2^2\; \frac{1\,-\,\cos k^x \,\cos k^y } {\wdc^2} \;,
\label{CKXKY}
\ee
\be
 D(\k)\;=\; q_1 \sin k^x\; q_2^2 \; \frac{\cos k^x - \cos k^y}{\wdc ^2} \;,
\ee
\be
 \wdc \;=\; \sqrt{1 -  q_1^2 \sin^2 k^x}\,.  \label{wdc}   
\ee

The coefficients $\omega^{(4)}_{\mu}(\k)$, $\mu = 1,\,2\,,3$, will only 
be needed in the determination of the coefficient $g_2$ of the fourth order 
term of $e^{(2)}_{LG}(Q_2)$ which will be discussed later. 
We will first concentrate on the determination of the zeroth order term, $e_2$, 
and of the coefficient $r_2$ of the second order term of $e^{(2)}_{LG}(Q_2)$. 
Under the assumption that $g_2$ is positive, this will provide us with an 
expression for the IC-DC phase boundary.
    
With the above  expressions for $\omega^{(\nu)}_{1} + \omega^{(\nu)}_{2}$ and  
$\omega^{(\nu)}_{3}$, $\nu = 0,\,1$, we obtain for the coefficients of the 
Landau Ginzburg energy from Eqs.~(\ref{EMFK}), (\ref{ELG})

\be
e_{2}(q_1, \lambda_a) \;=\; \frac{\la^2 \, q_1^2}{J} - 2 \la 
\left[1 + \kappa - \frac{1}{\ND} \sum_{\k} \wdc \right] \,,
\label{e02}
\ee

{\setlength{\arraycolsep}{-5pt}
\bea
&&\hspace*{-2mm}r_2(q_1,\, q_2,\, \la,\, |x_3(\kmin)|^2)\, = \NL{3mm}
&&2 -\,\frac{1}{\la}\, \frac{1}{q_2^2}\,(\kappa\,+\,1) \NL{3mm}
&&-\,\frac{1}{\lambda_a}\, \frac{1}{\ND} \sum_{\k}\, 
 \frac{1 - \cos k^x\,\cos k^y}{\wdc} \NL{3mm}
&& +\, \frac{1}{\ND} \left[\sum_{\k} \, \omega^{(2)}_3 (\k) 
+ |x_3(\kmin)|^2\, \omega^{(2)}_3 (\kmin) \right]\,.\label{r2}
\eea }

These are valid for arbitrary values of the parameters $q_1$, $\la$, $q_2$, 
and $|x_3(\kmin)|$.
In the next subsection, we will calculate their saddle point values for given 
$Q_2$ and thus fix the parameters. 
Here, we have only allowed for the existence of a condensate component 
$|x_3(\kmin)|^2$.
This is justified since, as Eqs.~(\ref{OMEXP2}) and (\ref{OMEXP}) show,
$\omega_3 < \omega_{1,2}$ for sufficiently small $Q_2$ so that 
conceivably $\omega_3(\k)$ may vanish at some point $\kmin$ in the the 
Brillouin zone, while $\omega_1(\k)$ and $\omega_2(\k)$ remain 
finite at $\kmin$, and hence a finite condensate density $|x_3(\kmin)|^2$ 
may occur at this point. 

\subsubsection{Saddle point, phase boundary}
Next we need to determine the saddle point of $e^{(2)}_{LG}(Q_2)$ in the space 
of the variables $q_1$, $\la$, $q_2$, and $|x_3(\kmin)|$. 
First, the saddle point values of  $q_1$ and $\la$ are obtained as expansions 
in powers of $Q_2$,
\begin{subequations}
\bea
\la &=& \la^{(0)} + \la^{(2)}\,Q_2^2 \, + \mathcal{O}(Q_2^4)\,, \label{LAMEX}\\
q_1 &=& q_1^{(0)} + q_1^{(2)}\,Q_2^2 \, + \mathcal{O}(Q_2^4)\,, \label{QEX}
\eea \label{QLEX}
\end{subequations}  
where $\la^{(0)}$, $q_1^{(0)}$ are the solutions of 
\begin{subequations}
\bea
\partial_{\la} e_{2} &=& 0\,, \label{DLAMe02} \\ 
\partial_{q_1} e_{2} &=& 0\,.  \label{DQe02}  
\eea \label{DLQe02}
\end{subequations}

Since the first derivatives of $e_{2}$ vanish at $\la = \la^{(0)} $, 
$q_1 = q_1^{(0)}$,  Eqs.~(\re{DLQe02}), we have 

\be
e_{2}\, = \,e_{2}^{(0)}\, + \,e_{2}^{(2)}\,Q_2^4\, + \mathcal{O}(Q_2^6),
\label{EXPe2}
\ee 

and

\be
r_2\, = \,r_2^{(0)}\, + \,r_2^{(1)}\,Q_2^2\, + \mathcal{O}(Q_2^4)
\label{EXPr2}
\ee
Here, $e_{2}^{(0)}$ and $r_2^{(0)}$ are the expressions (\re{e02}) and (\re{r2})
with $\la$ and $q_1$ replaced by $\la^{(0)}$ and $q_1^{(0)}$. 
The fourth order term of $e_{2}$, Eq.~(\re{EXPe2}), and the second order term 
of $r_2$ contribute only  to the fourth order term of $e^{(2)}_{LG}$ which 
will be determined later. 
Therefore, we postpone the presentation of explicit expressions for $\la^{(2)}$,
$q_1^{(2)}$ and the ensuing expressions for $e_{2}^{(2)}$ and $r_2^{(1)}$ until
later. 
With $e_2$ from Eq.~(\re{e02}), Eqs.~(\re{DLQe02}) yield the equations

\be
\kappa\,=\,\frac{2}{\pi} \mathbf{K}(q^{(0)}_1) -1,
\label{q1}
\ee

\be
\frac{\la^{(0)}}{J}\,=\,\frac{1}{(q^{(0)}_1)^{{}^{{}^{\scriptstyle 2}}}}
   \frac{2}{\pi} \left[  \mathbf{K}(q^{(0)}_1)\,-\,
\mathbf{E}(q^{(0)}_1) \right]\,,
\label{lama}
\ee 

which determine the saddle point values $q^{(0)}_1$ and $\la^{(0)}$.
( $\mathbf{K}$ and $\mathbf{E}$ are the elliptic integrals of the first and the 
second kind.)
 
Next we seek the extremum of $e^{(2)}_{LG}$ w.r.t.~$q_2$. 
Since $e_{2}$ is independent of $q_2$ we, neglecting terms of order 
$Q_2^4$, have 

\bea
0&=&\partial_{q_2}\,r^{(0)}_2 \NL{2mm}
&=&\frac{2}{q_2^3\,\lambda_a} \Bigg\{ \kappa\,+\,1\,-\,I_1(q_1,\,q_2)  \NL{2mm}
&&  - \frac{1}{\ND}\,\frac{C(\kmin)}{\lambda_a\,q^2_2}\,\frac{|x_3(\kmin)|^2}
   {\omega^{(2)}_3(\kmin)} \Bigg\}.  
 \label{q2}
\eea

and 
\be
I_1(q_1,\,q_2)\,=\,\frac{2}{\pi}\int_0^{\pi/2}\!\!\!dk^x\,\frac{1}{\pi}
\int_0^{\pi}\!\!\!dk^y\, \frac{C(\k)}{\left[ C(\k)^2 - D(\k)^2 \right]^{1/2}}.
\label{INT1}
\ee

(In these expressions and in the sequel, we use an abbreviated notation: 
$\la$, $q_1$ and $q_2$ denote the zeroth order quantities $\la^{(0)}$, 
$q^{(0)}_1$ and $q_2^{(0)}$.)
$\kmin$ is the location of the minimum of $\omega_3^{(2)}(\k)$,

\be 
k^y_{min}\,=\,0\,;\;\; 
  \left|\tan(\frac{k^x_{min}}{2})\right|\,=\,\frac{1}{q_1}\,, 
 -\pi\leq k^x_{min} \leq -\frac{\pi}{2}\,. \label{KMIN}
\ee

From (\re{OMEXP32}) and (\re{KMIN}) it follows that 
$\omega_3^{(2)}(\kmin)\,=\,0$, if 

\be
q_2^2\,=\,(1\,-\,q_1^2)/2.
\label{Q2Q1}
\ee
As a function of $q_2$ the integral $I_1(q_1,\,q_2)$ increases monotonously,

\bea
 &&1= I_1(q_1,\,0)\,\leq I_1(q_1,\,q_2) \leq I_1\left(q_1\right) \NL{3mm} 
 &&\mbox{for}\quad 0\,\leq\,q_2\,\leq\,\sqrt{(1\,-\,q_1^2)/2}\;. 
\label{VARINT}
\eea

We have defined
 
\be
I_1(q_1):= \max_{\{q_2\}}I_1(q_1,\,q_2)\,=\,
     I_1\left(q_1,\,\sqrt{(1\,-\,q_1^2)/2}\right)\;.   \label{I11}
\ee

\begin{figure}
\psfrag{q1}{\hspace{2cm}$q_1$}
\psfrag{qs}[][l]{$q_{1s}$}
\psfrag{1+kappa}[r][r]{$\kappa(q_1)+1$}
\psfrag{I1}[r][r]{$I_1(q_1)$}
\includegraphics[width=6cm]{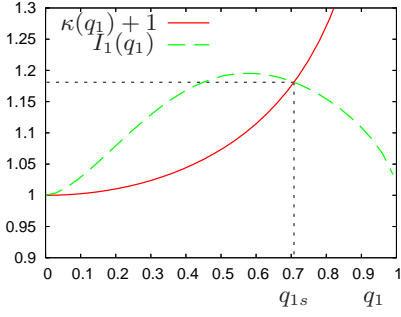}
\caption{\Co $I_1$ and $\kappa+1$ as functions of $q_1$.}
\label{IandK}
\end{figure}

As is seen in Fig.~\ref{IandK}, the graphs of the functions 
$\kappa \,= \kappa(q_1)$, 
Eq.~(\re{q1}), and of $I_1\,=\,I_1(q_1)$ intersect at $q_{1s} \simeq 0.708$, 
$\kappa_s \simeq 0.181$. 
Therefore, in solving Eq.~(\re{q2}) for $q_2$, two cases have to be considered 
separately:\\

\def\theparagraph{\roman{paragraph}}
\setcounter{paragraph}{0}
\paragraph{}
$q_1\,>\,q_{1s}$, $\kappa > \kappa_s$. 
In this case, a solution exists only, if the last term in parentheses in
Eq.~(\re{q2}) is positive. 
This requires that $\omega^{(2)}_3(\kmin) = 0$ because, as has been discussed 
before, $|x_3(\kmin)|$ and hence the ratio $|x_3(\kmin)|^2/\omega^{(2)}_3(\kmin)$
would vanish otherwise. The condition $\omega_3(\kmin)\,=\,0$ implies that 
$q_2^2\,=\,(1\,-\,q_1^2)/2$, cf. Eq.~(\re{Q2Q1}). Using this result and 
Eq.~(\re{lama}) to eliminate $q_2$ and $\lambda_a$ from Eq.~(\re{r2}) we find

\be 
r_2^{(0)} \;=\; 2 \left(1 - \frac{J_{DC}(\kappa)}{J} \right),
\label{R20}
\ee

where

\bea
J_{DC}(\kappa)&=&
\left[(\kappa + 1)\,(3 - q_1^2)/2 + 
  \tilde{I}_2(q_1, \sqrt{(1-q_1^2)/2}) \right]
\NL{1mm} && \cdot 
 \frac{2}{(1 -q_1^2)} \;\frac{q_1^2\,\pi}{4[\mathbf{K}(q_1) - \mathbf{E}(q_1)]} 
\label{PBID}
\eea
with 
\be
\tilde{I}_2(q_1, q_2)\;=\;
  \frac{2}{\pi^2}\int_0^{\frac{\pi}{2}}\!\!\!dk^x\,\int_0^{\pi}\!\!\!dk^y 
\left[  C(\k)^2 - D(\k)^2 \right]^{1/2}
\label{I2}
\ee  
is the IC-DC phase boundary for $\kappa > \kappa_s$, \ie in the region 
where the ratio $|x_3(\kmin)|^2/\omega^{(2)}_3(\kmin)$ is finite. 
According to the discussion at the end of section \re{results}, 
cf Eq.~(\re{CORDC}), this is the region where LRO prevails along the 
decoupled chains, cf.~Fig.~\re{phasediag}.\\
In the development leading to Eq.~(\re{PBID}) for the phase boundary, 
we have not needed the solution of Eq.~(\re{PBID}) explicitly, 
but we note it here for completeness: 

\bea 
\frac{1}{\ND}\,\frac{C(\kmin)}{\lambda_a\,q^2_2}\,\frac{|x_3(\kmin)|^2}
   {\omega^{(2)}_3(\kmin)}&=&\frac{1}{\ND}|x_3|^2 \frac{1}{2} 
\sqrt{\frac{\frac{1}{q_2^2}-\frac{2}{1+3q_1^2} }
           {\frac{1}{q_2^2}-\frac{2}{1-q_1^2}  }   } \NL{5mm}
&=& 1 + \kappa - I_1(q_1)\,>\,0. 
\label{X3}
\eea

These relations show that while $|x_3(\kmin)| = 0$, the ratio 
$|x_3(\kmin)|^2/\omega^{(2)}_3(\kmin)$ remains finite.\\

\paragraph{}

$q_1\,<\,q_{1s}$, $\kappa < \kappa_s$. In this case, we must have

\be
I_1(q_1,\,q_2))\,<\,I_1\left(q_1,\,\sqrt{(1\,-\,q_1^2)/2}\right),
\label{I1<}
\ee

(see Eq.~(\ref{I11})). Consequently $q_2^2 < (1 - q_1^2)/2$\; so that 
$\omega_3^{(2)}(\kmin) > 0$ and hence no condensate can develop, $|x_3|^2 = 0$.
Then, Eq.~(\re{q2}) yields the equation

\be
I_1(q_1,\,q_2)\,=\,1 + \kappa 
\label{q1q2}
\ee 

which replaces Eq.~(\re{Q2Q1}) and determines $q_2$ as a function of $q_1$, 
$q_2\,=\,q_2(q_1)$.  Then, proceeding as in case ({\it i}) one finds for the 
IC-DC phase boundary in the region $\kappa\,<\,\kappa_s$

\bea 
J_{DC}(\kappa)&=& 
\left[(\kappa + 1)(1 + q_2^2) + \tilde{I}_2(q_1,\,q_2) \right] \NL{1mm} &&
 \frac{1}{q_2^2}\;
 \frac{q_1^2\,\pi} {4\left[ \mathbf{K}(q_1) - \mathbf{E}(q_1) \right]}  \;.
\label{PBID0}
\eea 
 
Here, $q_1\,=\,q_1(\kappa)$ from Eq.~(\re{q1}) and 
 $q_2\,=\,q_2(\kappa)$ from Eq.~(\re{q1q2})(with $q_1\,=\,q_1(\kappa)$).

We note here that inside the DC phase, \ie for $J > J_{DC}(\kappa)$,
where $Q_2 = \lc = 0$, the saddle-point values of $q_1$ and $\la/J$ 
and hence of $Q_1$ are independent of $J$, cf.~Eqs.~(\re{q1}), (\re{lama}). 
Hence the graphs of $Q_1$ and $\la$ for $J < J_{DC}$ and for $J > J_{DC}$ 
join smoothly at $J = J_{DC}$, cf. Figs.~\re{Q1}, \re{lambda}. 
Furthermore, it follows from Eq.~(\re{X3}), that the ratio 
$(|x_3(\pm\kmin)|^2/\ND)/ (\lambda_a\,q^2_2 \omega^{(2)}_3(\kmin))$, which 
occurs in the amplitude of the spin-spin correlation function, 
cf.~Eq.~(\re{CORDCA}), is also independent of $J$ inside the DC phase and 
retains the value that it has attained at the IC-DC transition line.  
    
\subsubsection{Stability of the phase boundary}
In deriving the phase boundary from the condition $r_2^{(0)}\,=\,0$ we have 
tacitly assumed that the coefficient $g_2$ of the fourth order term in the 
LG expansion, Eq.~(\re{ELG}), is positive. 
In the remaining part of this appendix we will sketch the 
steps which lead to the conclusion that this is indeed the case. 

Expanding in the expression (\re{ELG}) for $e^{(2)}_{LG}$ the coefficients 
$e_2$ and $r_2$ w.r.t.~the second order contributions to $q_1$ and $\la$,
$q_1^{(2)}$ and $\la^{(2)}$, cf.~Eqs.~(\re{QLEX}) 
we obtain
\be 
e^{(2)}_{LG}\,= \,e^{(0)}_2 + r^{(0)}_2\, Q^2_2 + (g_2 + g'_2)\, Q^4_2 + 
\mathcal{O}(Q^6_2)\,,
\label{ELGEXS} 
\ee
where
\be 
  g_2\,=\,\frac{1}{\ND} \sum_\k 
  \left( \omega^{(4)}_1 + \omega^{(4)}_2 + \omega^{(4)}_3 \right)
\label{G}
\ee
is the contribution to the fourth order term of $e^{(2)}_{LG}$ that arises from 
the fourth order terms of the frequencies $\omega_{\mu}$
in the sum in Eq.~(\re{EMFK}) whereas the contribution to $e^{(2)}_{LG}$ 
of the expansion of $e_2$ and $r_2$ is
\bea 
g'_2 &=& \frac{1}{2} 
\left( q_1^{(2)}\;\la^{(2)}\right) \left( 
   \begin{array}{cc}
   \partial^2_{q_1}e_2|_{{}_0} &  
   \partial_{q_1}\partial_{\la}e_2|_{{}_0}\\[2mm]
   \partial_{\la}\partial_{q_1}e_2|_{{}_0} & 
   \partial^2_{\la}e_2|_{{}_0}
  \end{array} 
 \right)\left( 
 \begin{array}{l}  q_1^{(2)}\\[2mm] \la^{(2)} \end{array} \right)  \NL{2mm}
&&+ \left( q_1^{(2)}\;\la^{(2)}\right) 
\left(\begin{array}{l} \partial_{q_1}r_2|_{{}_0}\\[2mm]
\partial_{\la}r_2|_{{}_0} \end{array} \right )\,.
\label{G'}
\eea
(In Eq.~(\re{G'}) the notations $\partial^2_{q_1}e_2|_{{}_0}$ etc.~indicate 
that after the derivatives have been taken the variables $q_1$,  $\la$ 
etc.~have to be replaced by their zeroth order values $q^{(0)}_1$, $\la^{(0)}$ 
etc.)

The evaluation of the contribution (\re{G}) is straightfoward: 
the coefficients $\omega^{(4)}_{\mu}$, $\mu=1,\,2,\,3$, were obtained by 
solving Eq.~(\ref{DETD}) for $\omega$ iteratively to fourth order. 
As the explicit expressions are rather lengthy and contain no direct 
information, we refrain from presenting them here. 
The sum over $\k$ that is required in Eq.~(\re{G}) was done numerically.
$g_2$ was obtained in the form 
\be
g_2\,=\,\frac{1}{\la^3}\,\tilde{g_2}(q_1)\,,
\label{GQ1}
\ee
where $\tilde{g_2}(q_1)$ is a function of $q_1$ alone which is always 
positive so that $g_2\,>\,0$ throughout. 
Remarkably, no explicit dependence on the coupling constant $J$ appears in 
these results.

The evaluation of $g'$, Eq.~(\re{G'}) requires the knowledge of explicit 
expressions for $q_1^{(2)}$ and $\la^{(2)}$. 
These are obtained by expanding $e_2$ to first order in $q_1^{(2)}$ and 
$\la^{(2)}$, inserting the results into the expression (\re{ELG}) for 
$e^{(2)}_{LG}$ and requiring that the terms of order $Q_2^2$ satisfy the 
extremum conditions w.r.t.~$q_1$ and $\la$:

\begin{subequations}
\bea
0\,=\, q_1^{(2)}\partial^2_{q_1}e_2|_{{}_0} + 
 \lambda_a^{(2)}\partial_{q_1} \partial_{\lambda_a}e_2|_{{}_0} +
\partial_{q_1} r_2|_{{}_0}\,, \label{DQ12}\\[2mm]
0\,=\,  q_1^{(2)} \partial_{q_1}\partial_{\lambda_a} e_2|_{{}_0} +  
\lambda_a^{(2)}\partial^2_{\lambda_a} e_2|_{{}_0} + \partial_{\lambda_a} r_2|_{{}_0}\,.
\label{DLAM2}
\eea
\end{subequations}

The solution of these equations reads

\be
\left( \begin{array}{l} q_1^{(2)}\\ \lambda_a^{(2)} \end{array} \right)\,=
-\hat{M}\,\left(\begin{array}{l} \partial_{q_1} r_2|_{{}_0}\\ \partial_{\lambda_a} r_2|_{{}_0} 
\end{array} \right)\,, 
\label{SOLq2l2}
\ee

with
\be
\hat{M}^{-1}\,=\,\left( 
 \begin{array}{cc} \partial^2_{q_1}e_2|_{{}_0} & \;
                   \partial_{q_1}\partial_{\la}e_2|_{{}_0}\\[2mm]
                   \partial_{\la}\partial_{q_1}e_2|_{{}_0} & \;
                   \partial^2_{\la} e_2|_{{}_0}
 \end{array} \right)\,.  \label{MATR}
\ee

Inserting these results into Eq.~(\re{G'}) one finds 

\be
g'_2\,=\,-\frac{1}{2} \left( \begin{array}{l} \partial_{q_1} r_2|_{{}_0}\;\;
\partial_{\lambda_a} r_2|_{{}_0} \end{array} \right)\,\hat{M}\,
\left(\begin{array}{l} \partial_{q_1} r_2|_{{}_0}\\
\partial_{\lambda_a} r_2|_{{}_0}\end{array} \right)\,.
\label{G'G'}
\ee

While the second derivatives of $e_2$ are obtained straightforwardly from 
Eq.~(\re{e02}) the derivatives $\partial_{q_1} r_2|_{{}_0}$ and 
$\partial_{\la} r_2|_{{}_0}$ have to be calculated separately for the 
region $q_1 < q_{1s}$, where there is no condensate, $|x_3(\kmin)|^2 = 0$,
and for the region $q_1 > q_{1s}$, where $|x_3(\kmin)|^2 > 0$. 
Finally, the result for $g'$ can be cast into the form

\be   
g'_2\,= \frac{1}{\lambda_a^3}\,
     \left( \begin{array}{ll} x_q\; & \;x_{\lambda} \end{array} \right) \, 
     \hat{M}' \, 
     \left( \begin{array}{l} x_q \\ x_{\lambda} \end{array} \right) \,
\label{G'FIN}
\ee

where

\be
\hat{M}'\,=\,\frac{1}{4 q_1^2\, \Lambda (\kappa + 1 - \Lambda)} 
\left( 
  \begin{array}{cc} \frac{\displaystyle \Lambda}{\displaystyle 1-q_1^2} 
    & \; - \Lambda\\[3mm]
     -\Lambda & \; (2-q_1^2)\Lambda - \kappa - 1 \end{array} 
 \right)\, \label{SM}
\ee

with 

\be
\Lambda\,\equiv\,\frac{1}{q_1^2}\, \frac{2}{\pi} 
  \left[ \mathbf{K}(q_1) - \mathbf{E}(q_1) \right]
 \quad (\mathbf{K},\;\mathbf{E}\,:\;\; \mbox{elliptic integrals})
\ee

and 

\bea 
&& x_q\,=\, q_1^2 (\Lambda -\kappa -1) + 
 \left. \frac{1-q_1^2}{q_2^2} q_1 
\partial_{q_1}\tilde{I}_2(q_1,q_2)\right|_{q_2 = q_2(q_1)}  \NL{1mm} 
&& \hspace{1cm} -\Theta(q_1 -q_{1s}) \frac{4 q_1^2}{1 - q_1^2}
  (\kappa + 1 - I_1(q_1))\;, \NL{1mm} 
&&  x_{\lambda} \,=\, \left. ( \frac{1}{q_2^2} + 1 )(\kappa + 1) - 
 \frac{1}{q_2^2}\tilde{I}_2(q_1,q_2)\right|_{{}_{q_2 = q_2(q_1)}} .
\eea  

Here $\Theta$ is the step function; the integrals $I_1(q_1)$ and 
$\tilde{I}_2(q_1,q_2)$ have been defined above, cf.~Eqs.~(\re{I11}) and 
(\re{I2}), respectively. 
After numerical evaluation of these integrals, we find that 
$g_2' = g_2'(q_1)$ is positive for all values of $q_1$.

\bibliography{akago}
\end{document}